\documentclass[%
 onecolumn,
superscriptaddress,
 amsmath,amssymb,
 aps,
]{revtex4-2}

\usepackage[compact]{titlesec}
\titlespacing{\section}{0pt}{3ex}{3ex}
\titlespacing{\subsection}{0pt}{2ex}{2ex}
\titlespacing{\subsubsection}{0pt}{1ex}{1ex}

\usepackage{graphicx}% Include figure files
\usepackage{dcolumn}% Align table columns on decimal point
\usepackage{bm}% bold math
\usepackage[font=small,labelfont=bf,
   justification=Justified,
   format=plain,belowskip= 0 pt,aboveskip=0 pt]{caption} 
\usepackage{subcaption}
\usepackage[dvipsnames]{xcolor}
\usepackage{comment}
\usepackage{xfrac}
\newcommand\RD{\color{Blue}}

\begin{document}

\preprint{APS/123-QED}

\title{Rupture of a surfactant-laden draining thin film}% Force line breaks with \\

\author{Atul S Vivek}
 \email{me22resch04004@iith.ac.in}
\affiliation{Control Actuation Systems Group, Vikram Sarabhai Space Center, Indian Space Research Organisation, India\\
}%
\affiliation{Department of Mechanical \& Aerospace Engineering, Indian Institute of Technology Hyderabad, India\\
}%

\author{Ranabir Dey}
\email{ranabir@mae.iith.ac.in}
\affiliation{Department of Mechanical \& Aerospace Engineering, Indian Institute of Technology Hyderabad, India\\
}%

\author{Harish N Dixit}
\email{hdixit@mae.iith.ac.in}
\affiliation{Department of Mechanical \& Aerospace Engineering, Indian Institute of Technology Hyderabad, India\\
}%
\affiliation{Polymers \& Biosystems Engineering, Center for Interdisciplinary Programs, Indian Institute of Technology Hyderabad, India\\
}%

\date{\today}% It is always \today, today,

\begin{abstract}
Surfactant-laden thin liquid films overlaid on solid substrates are encountered in a variety of industrial and biological settings. As these films reach submicron thickness, they tend to become unstable owing to the influence of long-range dispersion forces. In the current study, we investigate how gravitational drainage affects the stability attributes of such thin liquid films. Using scaling arguments, we demonstrate that gravity and dispersion forces can exert their influence simultaneously over a wide range of film thicknesses. In the lubrication limit, we carry out linear stability analysis and nonlinear simulations to understand the evolution of draining thin films. Linear stability indicates the existence of two unstable modes and two cut-off wavenumbers, as opposed to a single unstable mode and a unique cut-off wavenumber observed in stationary films. It is also found that surfactant-laden flowing films are more stable than stationary films with surfactants as well as draining films with clean interfaces. The origin of stabilization is identified as the enhanced surfactant perturbations generated due to drainage. We demonstrate that films exhibiting intermediate levels of surfactant activity and significant drainage exhibit the lowest rates of disturbance growth, leading to extending the time of rupture.
\begin{description}
\item[Keywords]
Thin films, drainage, van der Waal's interaction, surfactants.\end{description}
\end{abstract}

\maketitle

%\tableofcontents

\section{\label{sec:introduction} Introduction}

Rupture of sub-micron-sized thin liquid films on solid substrates under the influence of van der Waals dispersion force has been extensively studied in the context of industrial coatings as well as biological systems including precorneal tear films, mucosal airway lining etc \cite{sharma1986lipids,deWit1994nonlinear,zhang2003slip, sharma1996instability,oron1997long,craster2009review,braun2012dynamics,kondic2020liquid}.  
\begin{comment}
{\RD{[Please give some comprehensive references including some review papers; note that you have already mentioned `extensively studied'. So, good references are expected]}}.
The free surface of these films are often overlaid with a monolayer of insoluble surface active agents. Ruckenstein \& Jain \cite{ruckenstein2018spontaneous} studied the evolution of thin films in the lubrication limit using the hydrodynamic stability theory. It was shown that spontaneous rupture of the film can occur under the action of long-range dispersion forces. Interfacial tension and presence of surface active impurities were shown to retard the disturbance growth rates \cite{sharma1986lipids, deWit1994nonlinear}.
\end{comment}
Classically, it has been shown that in the lubrication limit, van der Waals force can trigger hydrodynamic instabilites in thin liquid films ultimately leading to their rupture \cite{ruckenstein2018spontaneous}. However, interestingly, interfacial tension and the presence of surface active impurities (surfactants) can retard the growth rates of such van der Waals instabilities \cite{sharma1986lipids, deWit1994nonlinear}. 
This stabilizing effect of surfactants is attributed to the Marangoni convection, which arises from the redistribution of surfactants due to interfacial deformations \cite{sharma1986lipids, deWit1994nonlinear,zhang2003slip}. Moreover, viscoelastic interfacial stresses, arising from surface rheology in surfactant-laden films have also been demonstrated to exert a stabilizing influence on the dynamics of thin films \cite{chatzigiannakis2021thin, choudhury2020enhanced,cunha2023breakup}.  
Recently, the rupture of ultra-thin films was further studied by relaxing the constraints of lubrication theory and instead using the Stokes theory of fluid flow \cite{moreno2020stokes}. 
However, predictions for rupture time from both the Stokes flow model and the lubrication model are nearly identical, except in films of sub-nanometer thickness. 

Over the years, stability characteristics of gravity-driven flows in liquid films have been thoroughly explored \cite{yih1963stability, edmonstone2004flow, wei2005Marangoni, wei2005surfactantshear,craster2009review}. Yih \cite{yih1963stability}, in a pioneering study, showed that a liquid film flowing down an incline can develop instabilities and become unstable when a critical Reynolds number is exceeded.
Subsequently, the linear stability of a gravity-driven flowing film was also investigated in the presence of insoluble surfactants \cite{blyth2004surfactant}.  
In the presence of these insoluble surfactants, two distinct modes were identified: a classical Yih mode associated with interface deflections and a second Marangoni mode associated with spatial variations in surfactant concentration.
In essence, surfactants were found to have a stabilizing effect on the Yih mode by raising the critical Reynolds number for instability. 
However, at lower Reynolds number, the Marangoni mode was found to decay more slowly than the Yih mode indicating a reduction in stability. 
\begin{comment}
For more detailed information, interested readers are directed to a comprehensive review on the stability of thin liquid films compiled by Craster and Matar \cite{craster2009review}.     
\end{comment}

Despite the extensive scientific literature on thin film stability analysis, to date, no study has addressed the influence of gravitational drainage on the stability of surfactant-laden thin films when long-range dispersion forces are at play. 
The scarcity of literature in this direction may be due to the prevailing notion that dispersion forces and gravitational drainage operate at vastly different length scales, and are seldom encountered concomitantly. However, we argue to the contrary. We demonstrate through scaling arguments that simultaneous occurrence of both van der Waals interactions and gravitational drainage is practically feasible in thin films over a broad range of physical properties and  length scales. This observation holds relevance for thin films composed of various liquid materials, ranging from aqueous biological films such a pre-corneal tear films \cite{braun2012dynamics} to liquid metal films encountered during fabrication of plasmonic nanostructures \cite{makarov2016controllable}. 
Acknowledging the broad applicability of this compound problem,  all physical quantities and results presented in this paper are expressed in a dimensionless form to maintain generality. The specific case of an aqueous film is invoked in \S \ref{subsec:ScalingAnalysis} solely to establish the basic premise that van der Waals force and gravitational drainage can indeed be simultaneously relevant in physically realistic parameter regimes.

The rest of the paper is organized as follows. \S \ref{sec:MathForm}  gives the detailed mathematical formulation of the problem. We present a scaling analysis that establishes the relevant horizontal length scale for the system. The parameters regimes wherein gravitational drainage and van der Waal's dispersion forces become simultaneously relevant are also highlighted in this section. Subsequently, the linear stability of the system is investigated in \S \ref{sec:LSA}. Numerical setup for non-linear studies and the corresponding results are discussed in \S \ref{sec:NLA}. Concluding remarks and outlook are presented in \S \ref{sec:Conc}.

\begin{figure}[t]
\includegraphics[width=0.5\textwidth]{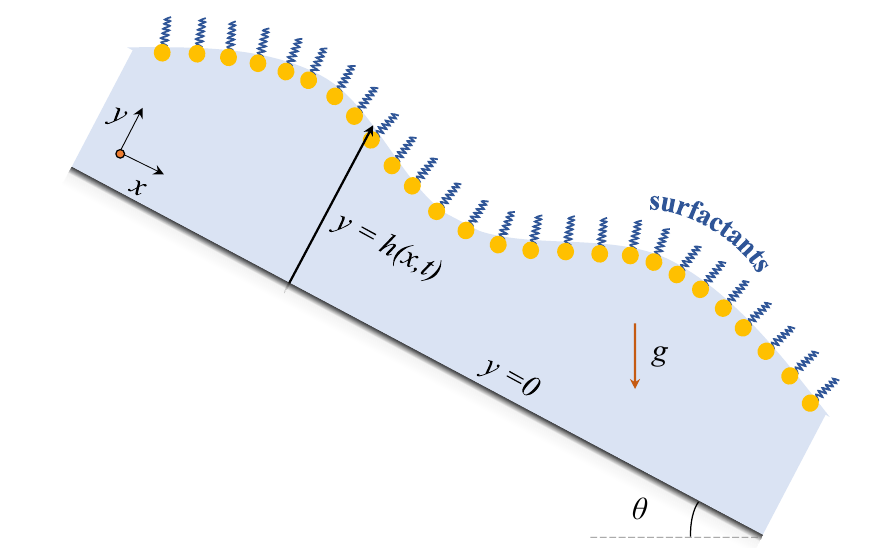}
\captionof{figure}{schematic of a thin liquid film flowing down an inclined plane, with surfactants overlaid on the film.}
\label{fig:Setup} 
\end{figure}

%%%%%%%%%%%%%%%%%%%%%%%%%%%%%%%%%%%%%%%%%%%%%
%%%%%%%%%%%%%%%%%%%%%%%%%%%%%%%%%%%%%%%%%%%%%
%%%%%%%%%%%%%%%%%%%%%%%%%%%%%%%%%%%%%%%%%%%%%
%%%%%%%%%%%%%%%%%%%%%%%%%%%%%%%%%%%%%%%%%%%%%
%%%%%%%%%%%%%%%%%%%%%%%%%%%%%%%%%%%%%%%%%%%%%
%%%%%%%%%%%%%%%%%%%%%%%%%%%%%%%%%%%%%%%%%%%%%
\section{\label{sec:MathForm} Mathematical Formulation}
We consider the stability of a thin film of Newtonian liquid laden with insoluble surfactant flowing down an impermeable inclined plane under the influence of gravity, as shown in Fig. \ref{fig:Setup}. The set-up is assumed to be isothermal and evaporation losses from the film are neglected. Incompressible Navier-Stokes governs the dynamics of the bulk fluid flow:
\begin{align}
\nabla\cdot\mathbf{u} &= 0 \label{Eq:Continuity} \\
\rho \left(\frac{\partial \mathbf{u}}{\partial t} + \mathbf{u} \cdot \nabla \mathbf{u}\right)& = -\nabla (p+\phi) + \nabla \cdot\boldsymbol \tau + \rho \mathbf{g}\label{Eq:NS} 
\end{align}
where $\mathbf u = \left[ u, v \right]^{\mathbf T}$  is the velocity vector,  $p$ is the pressure, $\phi = {A} /{6\pi h^3} $ is the disjoining pressure arising due to van der Waal's attraction \cite{ruckenstein2018spontaneous, zhang2003slip}, with ${A}$ denoting unretarded Hamaker constant, and $\mathbf{\tau}$ is the viscous stress tensor. For an angle of inclination $\theta$, the  acceleration due to gravity may be expressed in terms of its components along and normal to the plane as $\mathbf{g} = \left[g \sin(\theta), g \cos(\theta) \right]^{\mathbf T}$.
%
\begin{comment}
The deviatoric stress tensor,$\boldsymbol \tau$ is given by Eq.~(\ref{Eq:StressTensor}), wherein $\mu$ is the dynamic viscosity of the flowing liquid.
%
\begin{equation}\label{Eq:StressTensor} 
\boldsymbol \tau = \mu
\begin{bmatrix}
2  \dfrac{\partial u}{\partial x}     & \dfrac{\partial u}{\partial y} + \dfrac{\partial v}{\partial x}  \\
\dfrac{\partial u}{\partial y} + \dfrac{\partial v}{\partial x}    & 2  \dfrac{\partial v}{\partial y} \\
\end{bmatrix}
\end{equation}
%
\end{comment}
The evolution of the liquid-air interface $y=h(x,t)$ is governed by the kinematic condition\cite{zhang2003slip}:
\begin{equation} \label{Eq:kinCond}
\dfrac{\partial h}{\partial t} + u\dfrac{\partial h}{\partial x} = v
\end{equation}
The concentration of insoluble surfactants $\Gamma(x,t)$ at the interface follows the advection-diffusion equation \cite{leal2007advanced} given by
\begin{equation} \label{Eq:SurfactantTransport}
\dfrac{\partial \Gamma}{\partial t} + \mathbf{\nabla_s}\cdot({\mathbf u}\Gamma) = D_s \mathbf{\nabla_s}^2\Gamma
\end{equation}
Here, $\mathbf{\nabla_s} = \mathbf{\nabla} - \mathbf{n}(\mathbf{n}\cdot\mathbf{\nabla})$ is the surface gradient operator, with $\mathbf{n}$ being the unit normal vector to the interface, and $D_s$ denotes the surface diffusivity of the surfactants. Presently, we consider the surfactant concentration in the dilute limit. Hence, surface tension is expected to vary linearly with surfactant concentration as $\sigma =\sigma_0 - \Gamma({S}/{\Gamma_0}) $, where $\sigma_0$ is the surface tension of the clean interface, and S is the maximum spreading pressure given by $S = \sigma_0 - \sigma_s$. Here, $\sigma_s$ is the surface tension of the interface at the saturation surfactant concentration $\Gamma_0$ . Additionally, the normal and tangential stress balance conditions at the interface $y=h(x,t)$ may be prescribed as\cite{leal2007advanced}.
\begin{align}\label{Eq:NormStressBal}
-p +\mathbf{n}\cdot\mathbf{\tau}\cdot\mathbf{n} = -\sigma\mathbf{\nabla}\cdot\mathbf{n} \\
\label{Eq:TangStressBal}
\mathbf{n}\cdot\mathbf{\tau}\cdot\mathbf{t} = \mathbf{\nabla_s}\sigma\cdot\mathbf{t} 
\end{align}
where the $\mathbf{n}$ and $\mathbf{t}$ are the normal and tangential unit vectors to the  interface.
\begin{comment}
may be expressed in terms of the interface shape as: 
\begin{align}\label{Eq:UnitNorm}
\mathbf{n} = \frac{1}{\big(1+({\partial h}/{\partial x})^2\big)^{1/2}}\left[-{\partial h}/{\partial x}, 1\right]^{\mathbf T}  \\
\label{Eq:UnitTang}
\mathbf{t} = \frac{1}{\big(1+({\partial h}/{\partial x})^2\big)^{1/2}}\left[1, {\partial h}/{\partial x}\right]^{\mathbf T}
\end{align}
\end{comment}
%
Finally, the no-slip and impenetrability conditions on the wall at $y=0$ can be simply written as
\begin{align}\label{Eq:NoSlipNoPen}
u = 0, \ v = 0
\end{align}

%%%%%%%%%%%%%%%%%%%%%%%%%%%%%%%%%%%%%%%%%%%%%
%%%%%%%%%%%%%%%%%%%%%%%%%%%%%%%%%%%%%%%%%%%%%
%%%%%%%%%%%%%%%%%%%%%%%%%%%%%%%%%%%%%%%%%%%%%

\subsection{\label{subsec:ScalingAnalysis} Scaling Analysis}
Before proceeding with the non-dimensionalization of the governing equations, it is worthwhile to identify the major forces at play in the system and establish appropriate scales for the relevant physical variables.

In liquid films with a free surface, especially those that are sufficiently thin, attractive van der Waals force promotes the growth of perturbations arising at the interface. The physical mechanism of this instability can be easily explained in terms of a height-dependent disjoining pressure. When a wavy perturbation is imposed on an initially flat interface, the disjoining pressure is higher at the troughs of the perturbed film compared to the crests, driving fluid away from the troughs and amplifying the perturbation.
However, capillary forces at the interface and viscous forces oppose this growth of instability. 
Beyond a critical wavelength, dispersion forces prevail, ultimately leading to the rupture of the thin film \cite{ruckenstein2018spontaneous,sharma1985dryspots, deWit1994nonlinear, dey2019model,matar2002nonlinear}. 
It is possible to identify a wavelength with the maximum perturbation growth rate at which the rupture of the film is likely to occur \cite{zhang2003slip,matar2002nonlinear}. 

\begin{figure}[t]
\includegraphics[width=0.5\textwidth]{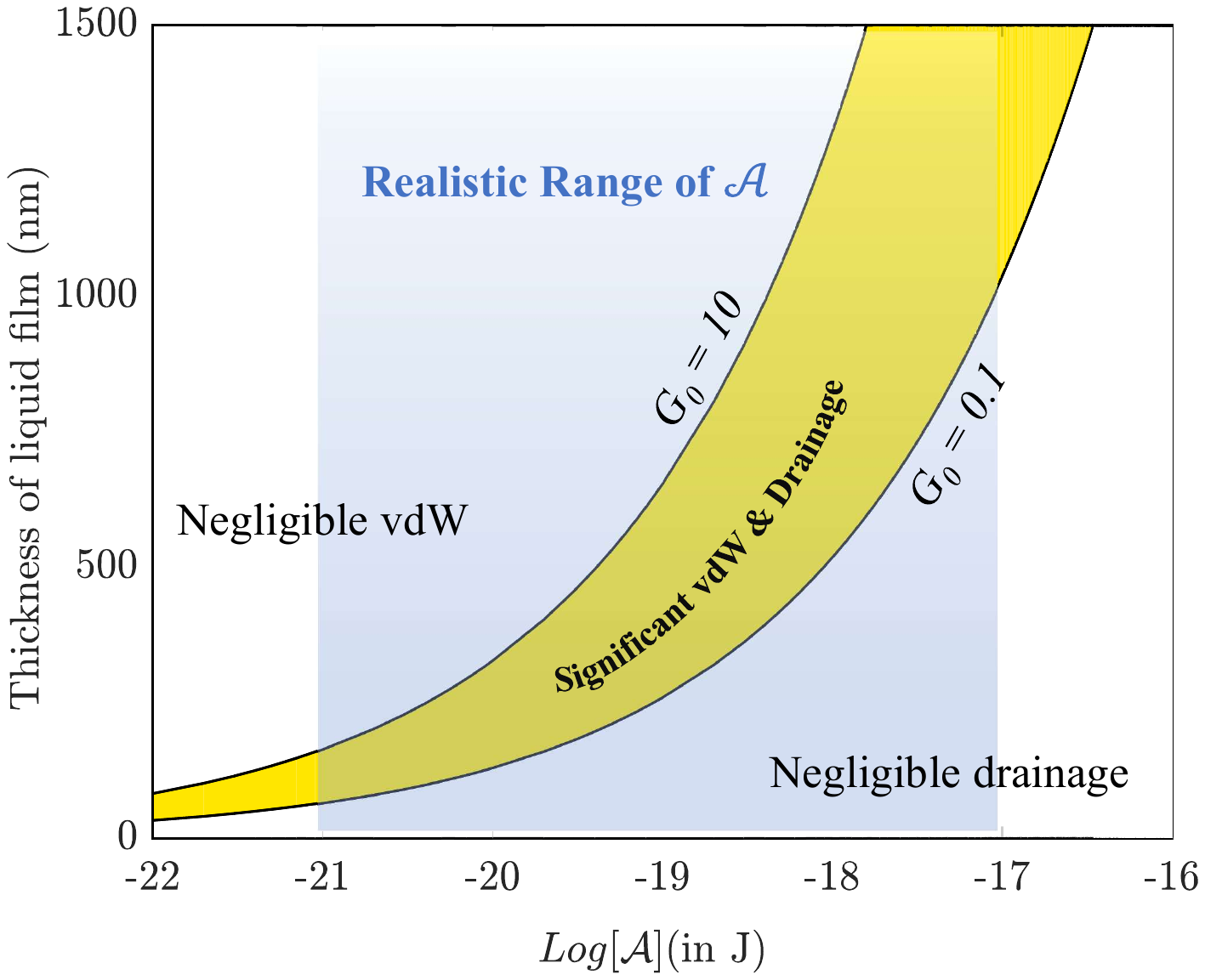}
\captionof{figure}{
The range of physical parameters for which both drainage and van der Waal's interaction are simultaneously relevant for an aqueous film ($ \rho = 1000 kgm^{-3}$, $g = 9.81 ms^{-2}$ and $\sigma_0 = 0.072 Jm^{-2}$). The yellow region bounded by the curves $G_0 = 0.1$ and $G_0 = 0.1$ indicates the theoretical parameter regimes for which gravitational and dispersion forces are significant. The translucent blue zone denotes the physically realistic range of Hamaker constants for an aqueous film. The intersection between the blue and yellow zones marks the physically realistic parameter regime for which both the aforesaid phenomena are significant.}
\label{fig:DrainageRelevant} 
\end{figure}

Regarding characteristic length scales in the problem, the unperturbed film thickness is an ideal choice for the scale for the film thickness.
For length scale along the thin film, the wavelength of the most unstable mode is the most appropriate scale, albeit its precise determination  necessitates a detailed stability analysis of the system. 
Nonetheless, a reasonable estimate can be obtained through scaling analysis.  
Previous studies on the subject \cite{deWit1994nonlinear, zhang2003slip} suggest that at very short perturbation wavelengths, surface tension dominates preventing the thin film from destabilizing. This condition entails that the dominant balance is between capillary and viscous forces. Conversely, for very long wavelength disturbances, van der Waal's dispersion forces can overcome capillary stabilization, resulting in a balance between viscous and dispersion forces. In this regime, the disturbance growth rates are extremely low. The most substantial growth is anticipated at intermediate wavelengths, where van der Waals dispersion forces and capillary forces tend to exhibit comparable magnitudes. This balance may be expressed in terms of scalings for the physical quantities as:
    \begin{equation}
    \frac{\mathcal{A}}{{H}^3L} \sim \frac{\sigma_0 H}{{L}^3}
    \end{equation}
    where $H$ is the thickness of the unperturbed film, $\sigma_0$ is the surface tension of the clean interface, $\mathcal{A}$ is the typical Hamaker constant and $L$ is the characteristic length scale along the thin film, which can be written as
    \begin{equation} \label{Eq:LengthScale}
    L = {H}^2\sqrt{\frac{\sigma_0}{\mathcal{A}}}
    \end{equation}
    To assess the validity of this estimate, a comparison is drawn with the linear stability results from \cite{deWit1994nonlinear}, for the special case of a surfactant-free thin film. In dimensional terms, the dispersion relation reduces to: 
     \begin{equation}\label{Eq:dewitDispersion}
     \omega  = - \frac{1}{3}\left(k^4 \frac{\sigma_0 {H}^3}{\mu} - k^2 \frac{3\mathcal{A}}{\mu H}\right)
    \end{equation}
    where $k$ is the wavenumber of the imposed perturbation. To obtain the peak growth rate, we set $\partial \omega/ \partial k = 0$ yielding $k_{max} = [ \mathcal{A}/(\sigma_0 {H}^4)]^{1/2}$. The corresponding scaling for the wavelength of the most unstable mode may be approximated as $\lambda \sim 1/k_{max} \sim {H}^2( \sigma_0/\mathcal{A})^{1/2}$, consistent with the scaling analysis. However, it's crucial to emphasize that this expression provides only an order of magnitude estimate. The precise value of the most unstable wavelength may be influenced by additional factors, such as surfactant transport and drainage, and necessitates a more rigorous analysis for exact determination. Nevertheless, as a length scale for significant fluid dynamics in the streamwise direction of the film, the aforementioned estimate holds appeal. 
   With a dominant length scale established for the rupture process, the relevance of gravitational drainage at this scale may be assessed. To this end, we consider a non-dimensional parameter $G_0 = \frac{\rho g \sin(\theta) {H}^3{L}}{\mathcal{A}} = \frac{\rho g \sin(\theta) {H}^5\sigma_0^{1/2}}{\mathcal{A}^{3/2}}$, which signifies the ratio of gravitational body forces to van der Waal's dispersion forces. Parameter regimes where $G_0$ is of $O(1)$ represent the physical conditions where these two forces are comparable. Let us consider the specific case of a vertically draining aqueous film with $ \rho = 1000 kgm^{-3}$, $g = 9.81 ms^{-2}$ and $\sigma_0 = 0.072 Jm^{-2}$. Depending on the nature of the underlying substrate, Hamaker constant for aqueous films may vary from $10^{-21}$ to $10^{-17}$ J. In the case of such a film, Fig. \ref{fig:DrainageRelevant} shows the range of film thicknesses where both van der Waals and gravitational body forces exhibit comparable magnitudes. We can thus conclude that gravity-driven drainage and van der Waals interaction may be simultaneously important over thicknesses spanning two orders of magnitude.  Notably, this range of film thicknesses encompasses the mucin layer in precorneal tear, along with airway linings \cite{zhang2003slip}, underscoring its relevance. 
   %{\RD{Important: Please refer to some physical problems (with citations) where this is important.}}  
   
\subsection{\label{SubSec:NonDimensionalization } Non-Dimensionalization of governing equations and thin-film approximation}

The system of equations can be non-dimensionalized as follows:
\begin{align*}
%\label{Scalings}
& x^* = \frac{x}{L}, \quad y^* = \frac{y}{H}, \quad h^* = \frac{h}{H}, \quad  u^* = \frac{u}{\textit U}, \quad v^* = \frac{Lv}{H \textit U},\\  t^* = &\frac{t \textit U}{L},  \quad p^* = \frac{p}{\textit P}, \quad A^* = \frac{A}{\mathcal{A}},\quad \Gamma^* = \frac{\Gamma}{\Gamma_{0}}, \quad \sigma^* = \frac{\sigma}{\sigma_{0}}, \quad S^* = \frac{S}{\sigma_{0}}
\end{align*}
where $\Gamma_0$ represents the maximum surfactant concentration on the interface. The characteristic streamwise perturbation velocity $U$ and characteristic pressure $P$ are defined as:
\begin{align}\label{Eq:Vel_Pres_Scaling}
\textit U = \frac{\cal A}{6 \pi \mu_{0} H L}, \qquad \textit P = \frac{\cal A}{6 \pi H^3},
\end{align}
These scaling choices are motivated by the dominant balance between van der Waals dispersion forces and viscous forces existing in the film. Following the discussion in \S\ref{subsec:ScalingAnalysis}, the characteristic streamwise length scale $L = H^2 \sqrt{\sigma_0/\mathcal{A}}$ is assumed to be similar in magnitude to the fastest growing rupturing mode in the film. These scalings are applied to Equns. (\ref{Eq:Continuity}) - ~(\ref{Eq:NoSlipNoPen}), resulting in the system of non-dimensional equations with boundary conditions given by Equns. ~(\ref{Eq:NonDim_Continuity}) to ~(\ref{Eq:NonDim_TangStressBal}). The asterisk $`*'$ symbol has been omitted from all non-dimensional quantities for convenience. 

\begin{align}
u_{x}+v_{y} &=0, \label{Eq:NonDim_Continuity}  \\
\epsilon^2  \textit {Re} \left(u_{t} +u u_{x} +v u_{y} \right) &= - (p_{x} +\phi_x) + \epsilon^2  u_{xx} + u_{yy} + G_0  , \label{NonDim_XMomentum}  \\
\epsilon^4  \textit {Re} \left(v_{t} + u v_{x} +v v_{y} \right) &= - p_{y} + \epsilon^4 v_{xx}   + \epsilon^2  v_{yy} + \epsilon G_0 \cot(\theta), \label{Eq:NonDim_YMomentum} 
\end{align}

Boundary conditions at the substrate, $y=0$ :
\begin{equation}
u = 0,  v = 0 .
\label{Eq:NonDim_NoSlipNoPen}
\end{equation}

Boundary conditions at the free surface, $y=h(x,t)$: \\
\indent \textit{Kinematic condition:}
\begin{equation}
h_t + uh_x = v, 
\label{Eq:NonDim_KinCond}
\end{equation}
\indent \textit{Surface transport of surfactants:}
\begin{equation}
\Gamma_{t} + \left(u_{s}  \Gamma \right)_{x} = \frac{1}{Pe} \left (\Gamma_{xx} \right), 
\label{Eq:NonDim_SurfactantTransport}
\end{equation}
\indent \textit{Normal stress balance condition:}
\begin{equation}
p  = \frac{2 \epsilon^2 }{1+\epsilon^2 h_{x}^2} \left[ \epsilon^2 u_{x} h_{x}^2 + v_{y} - \left(u_{y} + \epsilon^2 v_{x} \right) h_{x} \right] - \frac{{\cal \sigma} h_{xx}}{\left(1+\epsilon^2 h_{x}^2 \right)^{3/2}}, 
\label{Eq:NonDim_NormalStressBal}
\end{equation}
\indent \textit{Tangential stress balance condition:}
\begin{equation}
 \left [ \left( 1 - \epsilon^2  h_{x}^2 \right) \left(u_{y} + \epsilon^2 v_{x} \right) - 4 \epsilon^2 h_{x} u_{x} \right]= 
- \mathcal{M} \Gamma_{x} \left(1+\epsilon^2 h_{x}^2 \right)^{1/2}, 
\label{Eq:NonDim_TangStressBal}
\end{equation}
where $\epsilon = H/L$ is the ratio of the characteristic height and length scales, $ Re \left(= \rho U L/\mu_{0} \right)$ is the Reynolds number, $\phi$ is the non-dimensional van der Waal's potential,
%Subscripts $x$, $y$ and $t$ denote partial derivative with respect to the corresponding variables. 
$ Pe \left(= U L/D_{s} \right)$ is the P\'eclet number corresponding to the surfactant diffusion, and $ \mathcal{M} \left( =\epsilon \dfrac{S}{\mu_0 U} \right)$ denotes the ratio of Marangoni stresses to viscous stresses in the film. Further, the thickness of the film is considerably smaller than its streamwise length scale, and hence  $\epsilon \ll 1$. Using the lubrication approximation, the leading order equations, obtained by neglecting terms of $O(\epsilon)$, governing the evolution of the thin film can be written as
\begin{align}
 u_x + v_y&= 0, \label{Eq:Lubrication_Continuity} \\
  -(p+\phi)_x + G_0 + u_{yy} &=  0, \label{Eq:Lubrication_Xmomentum}\\
  -p_y  &= 0, \label{Eq:Lubrication_Ymomentum}\\  
\end{align}
with the following boundary conditions at $y =h(x,t)$:
\begin{align}
h_t+ uh_x &= v, \label{Eq:Lubrication_KinematicCond} \\
\Gamma_t +(\Gamma u)_x &= \frac{\Gamma_{xx}}{Pe}, \label{Eq:Lubrication_SurfactantTransport}\\
p &= - \sigma h_{xx},\label{Eq:Lubrication_NormalStressBal}\\ 
u_y& = -\mathcal{M} \Gamma_x,\label{Eq:Lubrication_TangentialStressBal}
\end{align}
and at $y=0$:
\begin{equation}
u=0, \quad v=0. \label{Eq:Lubrication_NoSlip_NoPen}\\ 
\end{equation}
From the momentum balance equation along the film thickness \big(Eq.~(\ref{Eq:Lubrication_Ymomentum})\big) and the normal stress balance at the interface \big(Eq. ~(\ref{Eq:Lubrication_NormalStressBal})\big), we obtain $p = p(x) = -\sigma h_{xx}$. 
Exact integration of the transverse momentum balance equation \big(Eq. ~(\ref{Eq:Lubrication_Xmomentum})\big), along with the tangential stress balance \big(Eq. ~(\ref{Eq:Lubrication_TangentialStressBal})\big) and the no slip boundary condition \big(Eq. ~(\ref{Eq:Lubrication_NoSlip_NoPen})\big), gives the streamwise velocity as, 

\begin{equation}
    u = \Big[ G_0 - \Big(-\sigma h_{xx} + \frac{A}{h^3}\Big)_x \Big] (hy-y^2/2) - \mathcal{M}\Gamma_x y
\end{equation}

\noindent Subsequently, using the continuity equation \big(Eq.~(\ref{Eq:Lubrication_Continuity})\big) and the no-penetration boundary condition \big(Eq. ~(\ref{Eq:Lubrication_NoSlip_NoPen})\big), we arrive at,
\begin{equation}
    v = \Big[ - G_0 + \Big(-\sigma h_{xx} + \frac{A}{h^3}\Big)_x \Big] h_x\frac{y^2}{2} + \Big(-\sigma h_{xx} + \frac{A}{h^3} \Big)_{xx}(h y^2/2- y^3/6) 
    + \mathcal{M}\Gamma_{xx} y^2/2
\end{equation}
Substituting the preceding expressions for fluid velocities into Eqs.~(\ref{Eq:Lubrication_KinematicCond}) and ~(\ref{Eq:Lubrication_SurfactantTransport}) for the  kinematic condition and surfactant transport respectively, we get the following system of coupled non-linear partial differential equations which governs the dynamics of the thin film:   

\begin{equation}\label{Eq:Thickness_Evolution}
h_t + G_0h_xh^2- \Bigg(\Big(\frac{A}{h^3} -\sigma h_{xx} \Big)_x \frac{h^3}{3}\Bigg)_x- \Big( \mathcal{M}\Gamma_x\frac{h^2}{2} \Big)_x = 0
\end{equation}

\begin{equation}\label{Eq:Surfactant_Evolution}
\Gamma_t  +  \Bigg( \Gamma\Big[ \frac{G_0h^2}{2}- \Big( \frac{A}{h^3} -\sigma h_{xx} \Big)_x \frac{h^2}{2} \Big] \Bigg)_x - \Big( \mathcal{M} \Gamma \Gamma_x h \Big)_x   = \frac{\Gamma_{xx}}{Pe} 
\end{equation}

\noindent  The stability of the above system of ODEs, with respect to perturbations in thickness and surfactant concentration, is studied in the subsequent sections. 

\section{\label{sec:LSA} Linear Stability Analysis}
To assess the linear stability characteristics of the system, we introduce normal mode perturbation of the form $h(x,t) =H_B + \tilde{h}e^{ikx+\omega t}$ \& $ \Gamma (x,t) =\Gamma_B + \tilde{\Gamma} e^{ikx+\omega t}$  into the evolution equations  ~(\ref{Eq:Thickness_Evolution}) and ~(\ref{Eq:Surfactant_Evolution}), where $h_B$ and $\Gamma_B$ represent the uniform thickness and surfactant concentration of the unperturbed film. Furthermore, $k$ denotes the Non-dimensional wavenumber of the perturbation, and $\omega$ is the perturbation growth rate corresponding to a wavenumber. The disturbances are assumed to be much smaller than the base state quantities, and hence, only the linear terms in perturbations are retained for evaluating the stability of the film. This yields the following homogeneous system of linear equations in  $\tilde{h}$ and $\tilde{\Gamma}$ : 
\begin{equation} \label{Eq:LSAEigenValueProblem}
\begin{bmatrix}
 \omega +ikG_0 H_B^2 -\Big[\frac{3Ak^2}{H_B^4}-\sigma k^4 \Big]\frac{H_B^3}{3} & \frac{\mathcal{M}k^2H_B^2}{2} \\
ikG_0 \Gamma_B H_B -\Big[\frac{3Ak^2}{H_B^4}-\sigma k^4 \Big]\frac{H_B^2 \Gamma_B}{2} & \omega + \frac{ikG_0 H_B^2}{2} + \mathcal{M}k^2 \Gamma_B H_B +\frac{k^2}{Pe} 
\end{bmatrix}  
\begin{bmatrix}
    \tilde{h} \\ \tilde{\Gamma}
\end{bmatrix}
=
\begin{bmatrix}
    0 \\ 0
\end{bmatrix}
\end{equation}

\noindent For non trivial solutions, the determinant of the coefficient matrix must be equal to $0$. Using this condition, the dispersion relation can be obtained as:  

\begin{multline}
\label{DispersionRelation}
\omega^2+ \omega\Bigg( \frac{3ikG_0 H_B^2}{2} - \Big[\frac{3Ak^2}{H_B^4}-\sigma k^4 \Big]\frac{ H_B^3}{3} + \mathcal{M}k^2 \Gamma_B H_B + \frac{k^2}{Pe} \Bigg) 
 + ikG_0H_B^2 \Bigg(\frac{ikG_0 H_B^2}{2}   + \frac{\mathcal{M}k^2 \Gamma_B H_B}{2} + \frac{k^2}{Pe} \Bigg) \\ - \Big[\frac{3Ak^2}{H_B^4}-\sigma k^4 \Big]\frac{H_B^3}{3} \Bigg(\frac{ikG_0 H_B^2}{2} + \frac{\mathcal{M}k^2 \Gamma_B H_B}{4} + \frac{k^2}{Pe}\Bigg) =0 
\end{multline}

\begin{figure}
     \centering
     \begin{subfigure}[b]{0.4\textwidth}
         \centering
         \includegraphics[width=\textwidth]{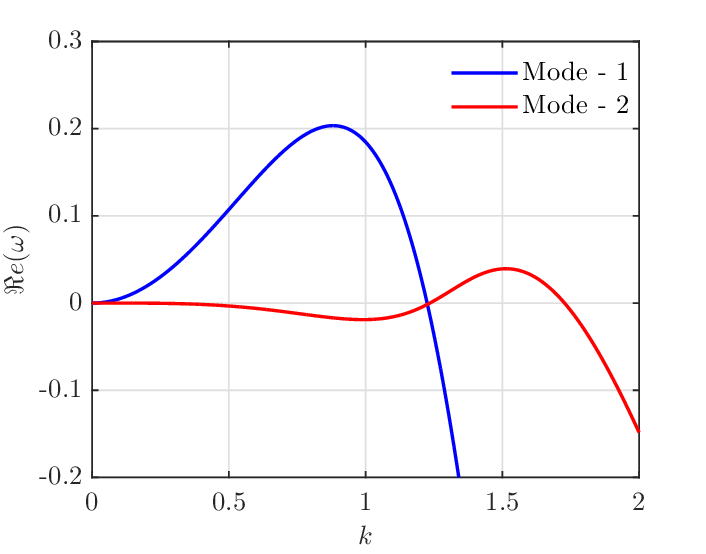}
         \caption{}
         \label{fig:GrowthRatePlot_Linear}
     \end{subfigure}
     \begin{subfigure}[b]{0.4\textwidth}
         \centering
         \includegraphics[width=\textwidth]{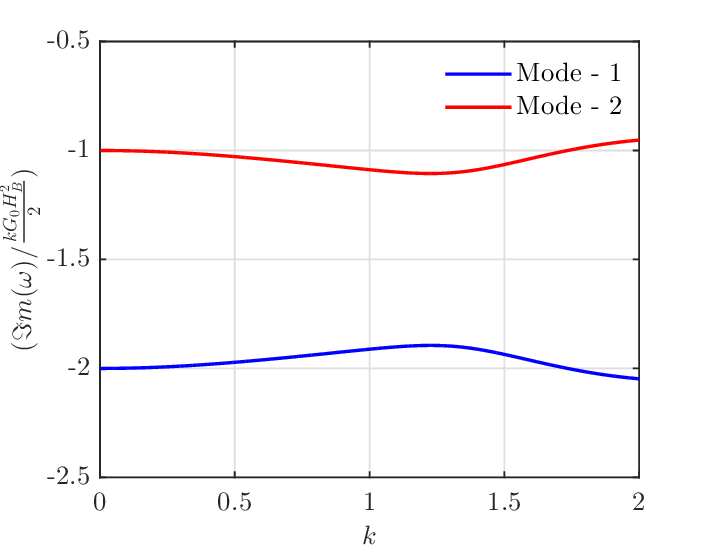}
         \caption{}
         \label{fig:WaveSpeedPlot_Linear}
     \end{subfigure}
        \caption{Dispersion relation for a surfactant-laden draining film rupturing under van der Waal's forces.(a) Non-dimensional perturbation growth rates. (b) Non-dimensional wave speeds for two modes normalized by the surface speed of the unperturbed draning film. Parameter values  are $G_0 = 2, \Gamma_B = 0.5, H_B=1, \sigma = 1, \mathcal{M} = 1,A=1, Pe=1000 $.} 
        \label{fig:DispersionRelationGeneric}
\end{figure}

\noindent As the dispersion relation is quadratic, for a definite wavenumber $k$, two values of $\omega$ are obtained which correspond to two different modes. The film is stable (unstable) to perturbation of a given wavenumber $k$, if the real part of the growth rate $\Big(\mathfrak{Re}(\omega) \Big)$ is negative (positive). The imaginary component of $\omega$ is associated with the speed at which the disturbances travel on the surface of the film. The solution of the dispersion relation for a set of parameters representing a film for which both gravitational drainage and van der Waals interaction are important is shown in Fig. \ref{fig:DispersionRelationGeneric}. The dispersion relation depicts two unstable eigenmodes for the film traveling at two different wave speeds, which in turn vary with the perturbation wavenumber. Specifically, Mode-1 travels at twice the surface speed whereas Mode-2 travels at the same speed as the surface of the draining film (Fig. \ref{fig:DispersionRelationGeneric}(b)). To physically identify these modes, we consider the special case of a flowing film laden with a passive insoluble species instead of surfactants ($\mathcal{M} = 0$). In this limit, the stability problem reduces to the following system of linear equations: 

\begin{equation}
\label{Eq:PassiveSurfactantCase}
\begin{bmatrix}
 \omega +ikG_0 H_B^2 -\Big[\frac{3Ak^2}{H_B^4}-\sigma k^4 \Big]\frac{H_B^3}{3} & 0 \\
ikG_0 \Gamma_B H_B -\Big[\frac{3Ak^2}{H_B^4}-\sigma k^4 \Big]\frac{H_B^2 \Gamma_B}{2} & \omega + \frac{ikG_0 H_B^2}{2} +\frac{k^2}{Pe} 
\end{bmatrix}  
\begin{bmatrix}
    \tilde{h} \\ \tilde{\Gamma}
\end{bmatrix}
=
\begin{bmatrix}
    0 \\ 0
\end{bmatrix}
\end{equation}

\begin{comment}
    
\noindent This stability problem also yields two eigenvalues and their respective eigenfunctions. The first mode given by Eq. ~(\ref{Eq:InterfaceModeGrowthRate}) and ~(\ref{Eq:InterfaceModeShape}) can be considered as an \textit{interface} mode, characterized by interfacial perturbation accompanied by corresponding disturbances in the species concentration. This unstable mode has perturbations traveling with wave speeds ($c = \mathfrak{Im}(\omega)/k = -G_0H_B^2$) twice that of the film's surface speed. It may be seen that this mode is similar to the Mode-1in Fig. \ref{fig:DispersionRelationGeneric} for the surfactant-laden draining film. Furthermore, it is worth noting that the perturbation growth rates for this mode is identical to that for a classical surfactant-free stationary film \cite{deWit1994nonlinear}. Hence, we can conclude that that drainage has little impact on the stability characteristics of the film without the surfactants.
\end{comment}      

\noindent This stability problem also yields two eigenvalues and their respective eigenfunctions. The first mode given by Eq. ~(\ref{Eq:InterfaceModeGrowthRate}) and ~(\ref{Eq:InterfaceModeShape}) can be considered as an \textit{interface} mode, characterized by interfacial perturbation accompanied by corresponding disturbances in the species concentration. This unstable mode has perturbation growth rates identical to that for a classical surfactant free stationary film \cite{deWit1994nonlinear} and remains unstable at very low wavenumbers.  Furthermore, the perturbations of this mode travel with wave speeds ($c = \mathfrak{Im}(\omega)/k = -G_0H_B^2$) twice that of the film's surface speed. Therefore, it may be seen that this mode is similar to the Mode-1 in Fig. \ref{fig:DispersionRelationGeneric} for the surfactant-laden draining film. Based on the former observation on the perturbation growth rate, we also can conclude that that drainage has little impact on the stability characteristics of the film without the surfactants. 

\begin{gather}
 \omega =  -ikG_0 H_B^2 +\Big[\frac{3Ak^2}{H_B^4}-\sigma k^4 \Big]\frac{H_B^3}{3} \label{Eq:InterfaceModeGrowthRate} \\
  \tilde{\Gamma} \Big[ -\frac{ikG_0 H_B^2}{2} + \Big[\frac{3Ak^2}{H_B^4}-\sigma k^4 \Big]\frac{H_B^3}{3} + \frac{k^2}{Pe} \Big] = \tilde{h} \Big[ ikG_0 \Gamma_B H_B -\Big[\frac{3Ak^2}{H_B^4}-\sigma k^4 \Big]\frac{H_B^2 \Gamma_B}{2} \Big] 
  \label{Eq:InterfaceModeShape} 
\end{gather}

\noindent The second mode for this special problem without any surfactants, described by Eq.~(\ref{Eq:SurfactantModeGrowthRate}) and ~(\ref{Eq:SurfactantModeShape}), denotes a \textit{species} mode triggered by a perturbation in the concentration of the insoluble species alone. This mode moves with nearly the same speed as the surface speed of the film ($c = \mathfrak{Im}(\omega)/k = -G_0H_B^2/2$) and is analogous to mode 2 in Fig. \ref{fig:DispersionRelationGeneric}. Note that when the passive species is replaced by a surface active agent, like surfactants, any perturbation in the species concentration generates a corresponding interfacial deflection through the Marangoni stresses. Further, it is worth noting that the observations outlined here are consistent with the studies on instabilities in surfactant-laden flowing films \cite{blyth2004surfactant,wei2005Marangoni}, wherein two distinct modes affecting the stability of the film were identified.
\begin{gather}
 \omega =  -\frac{ikG_0 H_B^2}{2} - \frac{k^2}{Pe} \label{Eq:SurfactantModeGrowthRate}, \\
  \tilde{h} = 0 \label{Eq:SurfactantModeShape}
\end{gather}

The dispersion relation for the present problem stands out in comparison to the dispersion relation for a thin film with insoluble surfactants, but without gravitational drainage, in a few key aspects. 
Importantly, the draining film presents two unstable eigenmodes, whereas stationary films exhibit a single unstable mode. 
A second important distinction from its stationary counterpart is the existence of  cut-off wavenumbers at which either of the two modes are neutrally stable (Fig. \ref{fig:DispersionRelationGeneric}(b)).
In addition, an exchange of instability occurs at the first or the lower cutoff wavenumber. This is in contrast to stationary films having a single cut-off wavenumber given by $k_c = \sqrt{3A/H_B^4\sigma}$. 
To obtain analytical expressions for the cut-off wavenumbers in the present case, we consider the limit of negligible surfactant diffusion ($Pe \to \infty$). Furthermore,  we rewrite $\omega = \omega_r+ i\omega_i$ in Eq.~( \ref{DispersionRelation}) and impose the conditions $\omega_r = 0 $ and
$ \omega_i \in \mathbb{R}$. Straightforward substitution and decomposition yields two cut-off wavenumber - an upper cut-off wave-number($k_{c,high}$) and a lower cut-off wavenumber ($k_{c,low}$) as:     
\begin{gather} 
    k_{c,low} = \sqrt{\frac{3A}{\sigma H_B^4}- \frac{3\mathcal{M}\Gamma_B}{\sigma H_B^2}} \label{Eq:LowerCutoffWaveNumber}\\
    k_{c,high} = \sqrt{\frac{3A}{H_B^4\sigma}} \label{Eq:UpperCutoffWaveNumber}
\end{gather}

\noindent The  latter solution is admissible if and only if the following condition is satisfied: 

\begin{equation} \label{Eq:LowerCutoffWavenumberCriteria}
    0 < \frac{3A}{\sigma H_B^4}- \frac{3\mathcal{M}\Gamma_B}{\sigma H_B^2} \leq \Bigg(\frac{G_0H_B}{2\mathcal{M}\Gamma_B}\Bigg)^2
\end{equation}

Going forward, it may also be worthwhile to discuss a few other features of interest pertaining to these cut-off wavenumbers. The upper cutoff wavenumber is identical to that in a stationary surfactant-laden film and occurs when $\frac{3Ak^2}{H_B^4}-\sigma k^4 =0$, denoting a balance between the destabilizing van der Waal's forces and the stabilizing capillary effects. However, the second cutoff wavenumber occurs when $-\Big[\frac{3Ak^2}{H_B^4}-\sigma k^4 \Big]\frac{H_B^3}{3} + \mathcal{M}k^2 \Gamma_B H_B=0$, which represents a special balance between van der Waal’s, capillary and Marangoni forces. It is important to emphasize that this equilibrium is observed only in the presence of surfactants and when drainage levels exceed a a critical threshold.

\subsection{Effect of gravitational drainage on stability \label{subsec:EffectofG0}}

\begin{figure}[t]
     \centering
     \begin{subfigure}[b]{0.42\textwidth}
         \centering
         \includegraphics[width=\textwidth]{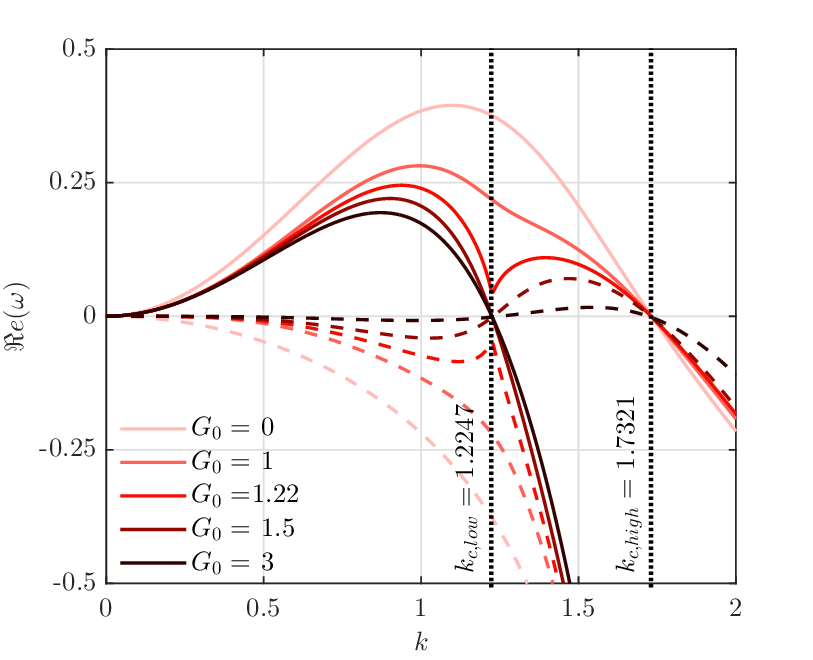}
         \caption{}
         \label{fig:EffG0_GrowthRate_Comp}
     \end{subfigure}
     \begin{subfigure}[b]{0.42\textwidth}
         \centering
         \includegraphics[width=\textwidth]{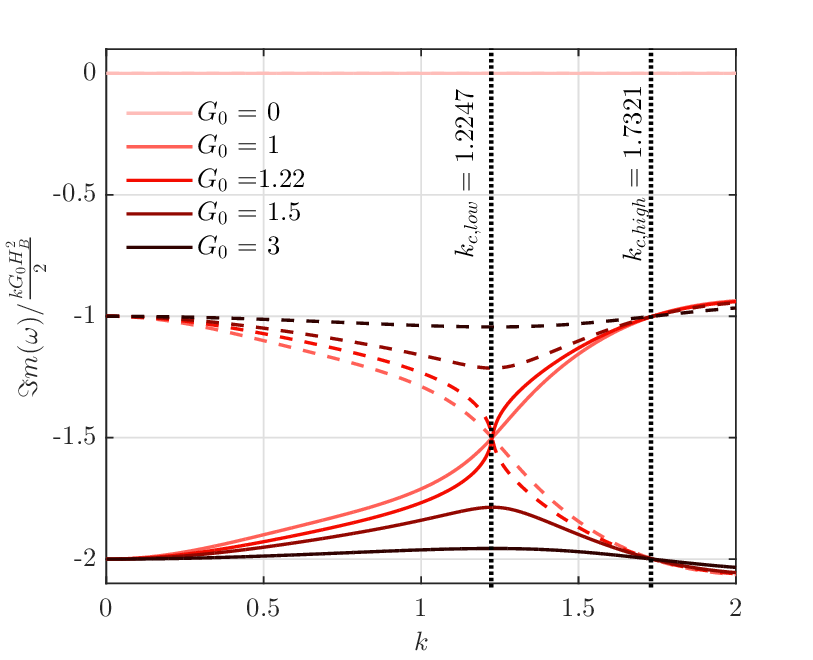}
         \caption{}
         \label{fig:EffG0_Wavespeed_Comp}
     \end{subfigure}
          \begin{subfigure}[b]{0.42\textwidth}
         \centering
         \includegraphics[width=\textwidth]{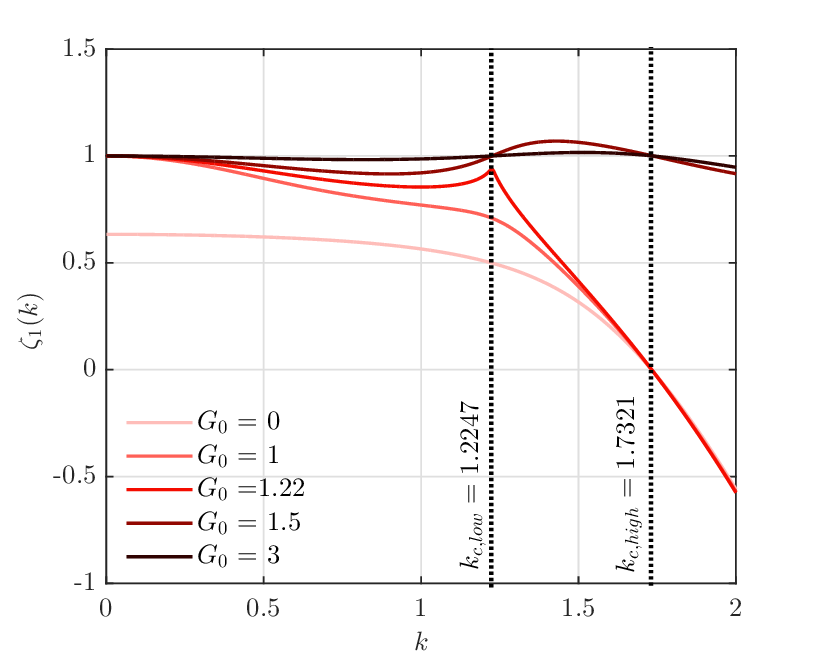}
         \caption{}
         \label{fig:EffG0_Projection_Mode1}
     \end{subfigure}
     \begin{subfigure}[b]{0.42\textwidth}
         \centering
         \includegraphics[width=\textwidth]{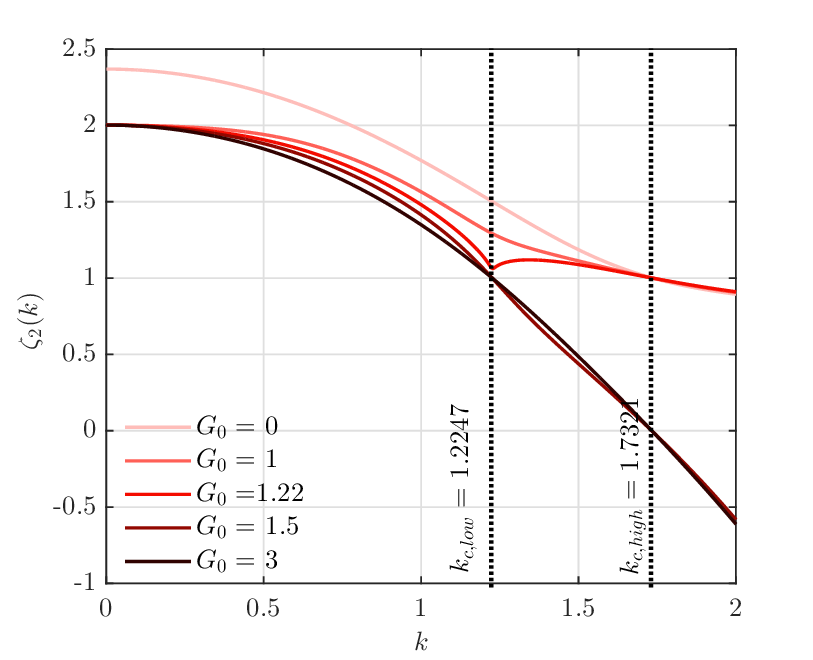}
         \caption{}
         \label{fig:EffG0_Projection_Mode2}
     \end{subfigure}
     
        \caption{ Effect of $G_0$ on stability of the film for the parameters:  $\Gamma_B = 0.5, H_B=1, \sigma = 1, \mathcal{M} = 1, A=1, Pe=1000 $.  Dispersion relations for different values of $G_0$ is presented. (a) Non-dimensional perturbation growth rates and (b) normalized perturbation wave speeds. Solid lines  and dashed lines in (a) and(b) denote interface mode (\textit{mode 1}) and surfactant mode (\textit{mode 2}) respectively. Projection function $\zeta_{1}(k)$ and  $\zeta_{2}(k)$ associated with the interface mode and surfactant mode are plotted in (c) and(d) respectively. The dotted vertical lines represent upper and lower cutoff wavenumbers for this parameter set.}
        \label{fig:EffectOfG0}
\end{figure}

The impact of the drainage parameter $G_0$ on the stability of the film is demonstrated in Fig. \ref{fig:EffectOfG0}, for a typical parameter set.
A cursory glance on the dispersion relation in Fig. \ref{fig:EffG0_GrowthRate_Comp} shows that increasing $G_0$ results in lower disturbance growth rates, suggesting that drainage apparently has a stabilizing effect. However, a more thorough examination reveals many additional features of interest. 
For the chosen set of parameters, the lower and upper cut off wavenumbers are $k_{c,low} = 1.225$ and $k_{c,high} = 1.732$ respectively. 
Note that criterion for existence of the lower cut off wavenumber, as given in Eq.~( \ref{Eq:LowerCutoffWavenumberCriteria}), is satisfied only for $G_0> 1.225$. 
For $G_0$ values below this limit the Mode-2 is always stable, and instability arises from the Mode-1 alone. 
As the limit $G_0 = 1.225$ is approached, a cusp emerges in the dispersion relation at the lower cutoff wavenumber. Near this point, the growth rate of the interface mode is suppressed while the growth rate of the surfactant mode is enhanced. 
At even higher values of $G_0$, both the modes become neutrally stable at $k_{c,low}$. 
Beyond the lower cutoff wavenumber, the surfactant mode emerges as the dominant unstable mode, indicating an exchange of instabilities.
For very rapid drainage (e.g., $G_0 = 3$), both the surfactant and interface modes experience attenuation.

\begin{figure}[t]
     \begin{subfigure}[h]{0.45\textwidth}
         \centering
         \includegraphics[width=\textwidth]{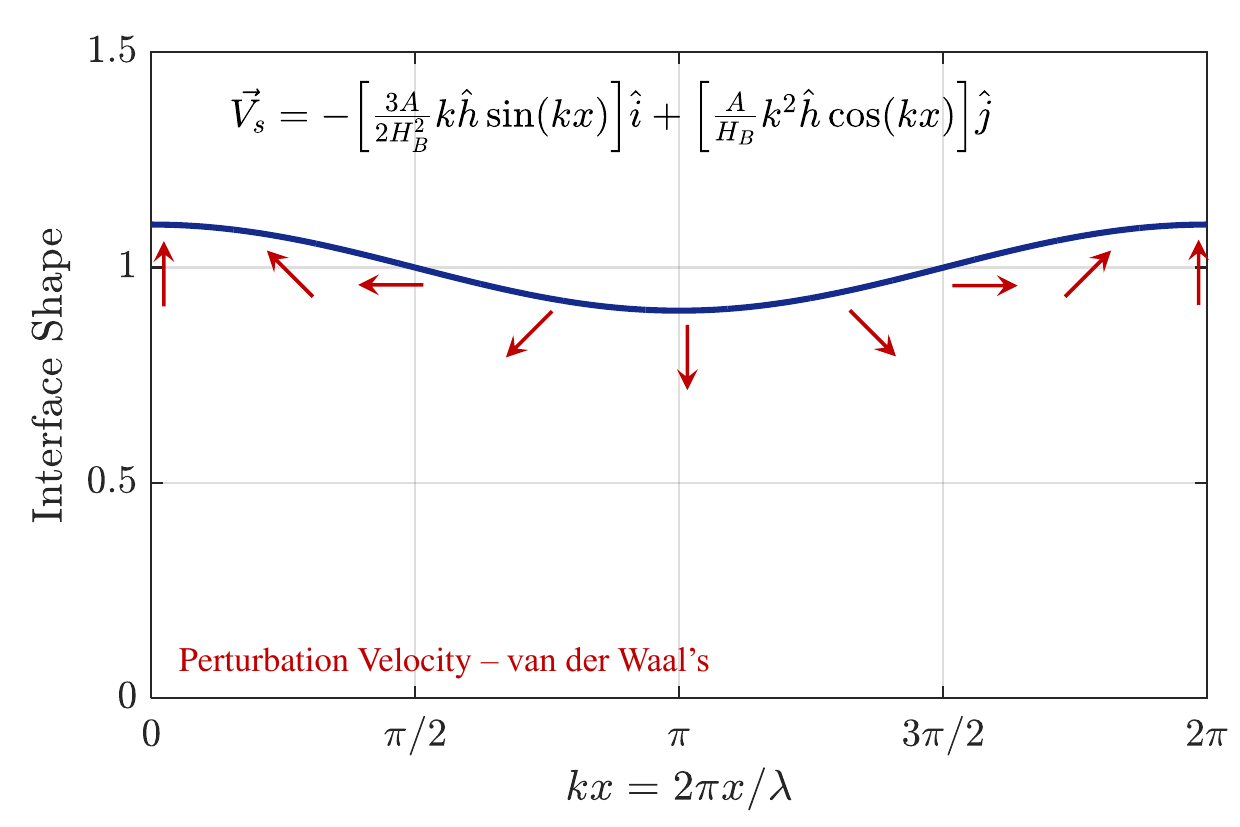}
         \caption{}
         \label{fig:VelocityFields_vdW}
     \end{subfigure}
     \begin{subfigure}[h]{0.45\textwidth}
         \centering
         \includegraphics[width=\textwidth]{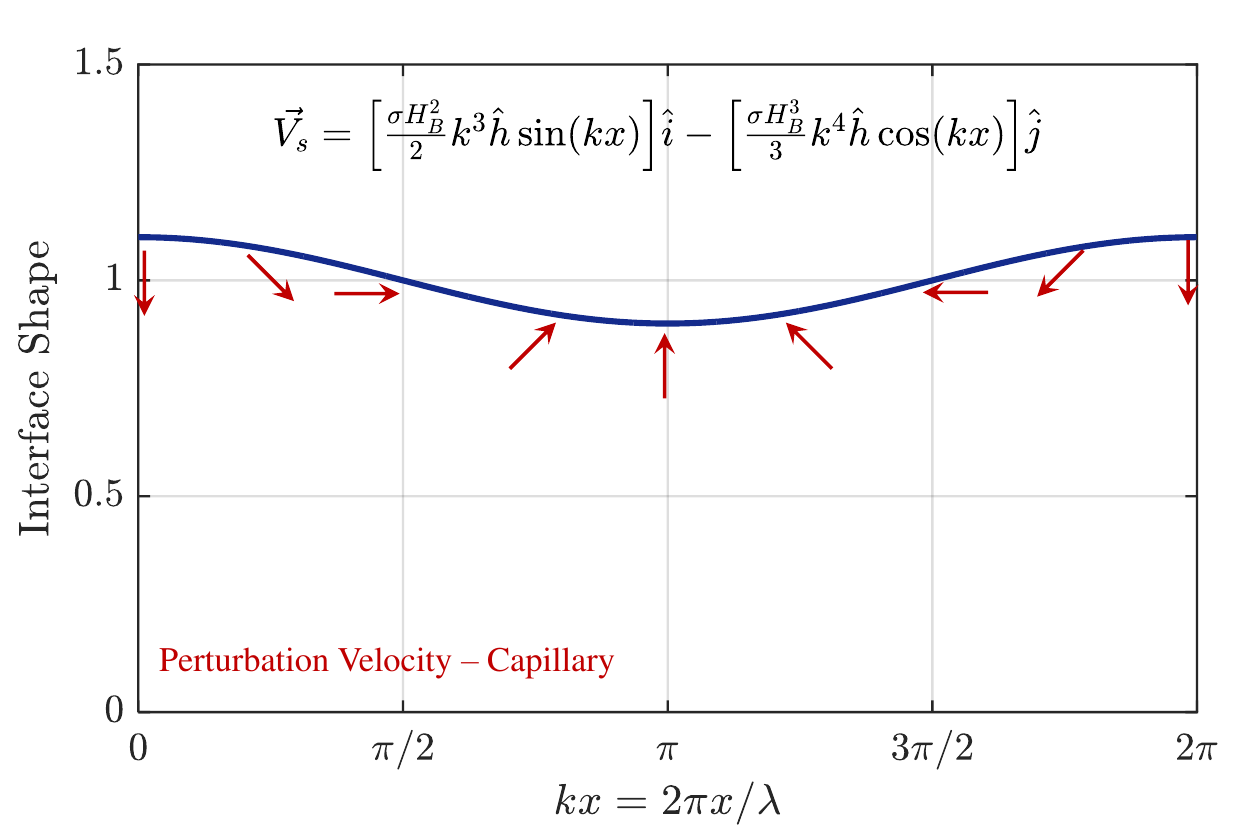}
         \caption{}
         \label{fig:VelocityFields_Capillary}
     \end{subfigure}
     \begin{subfigure}[h]{0.45\textwidth}
         \centering
         \includegraphics[width=\textwidth]{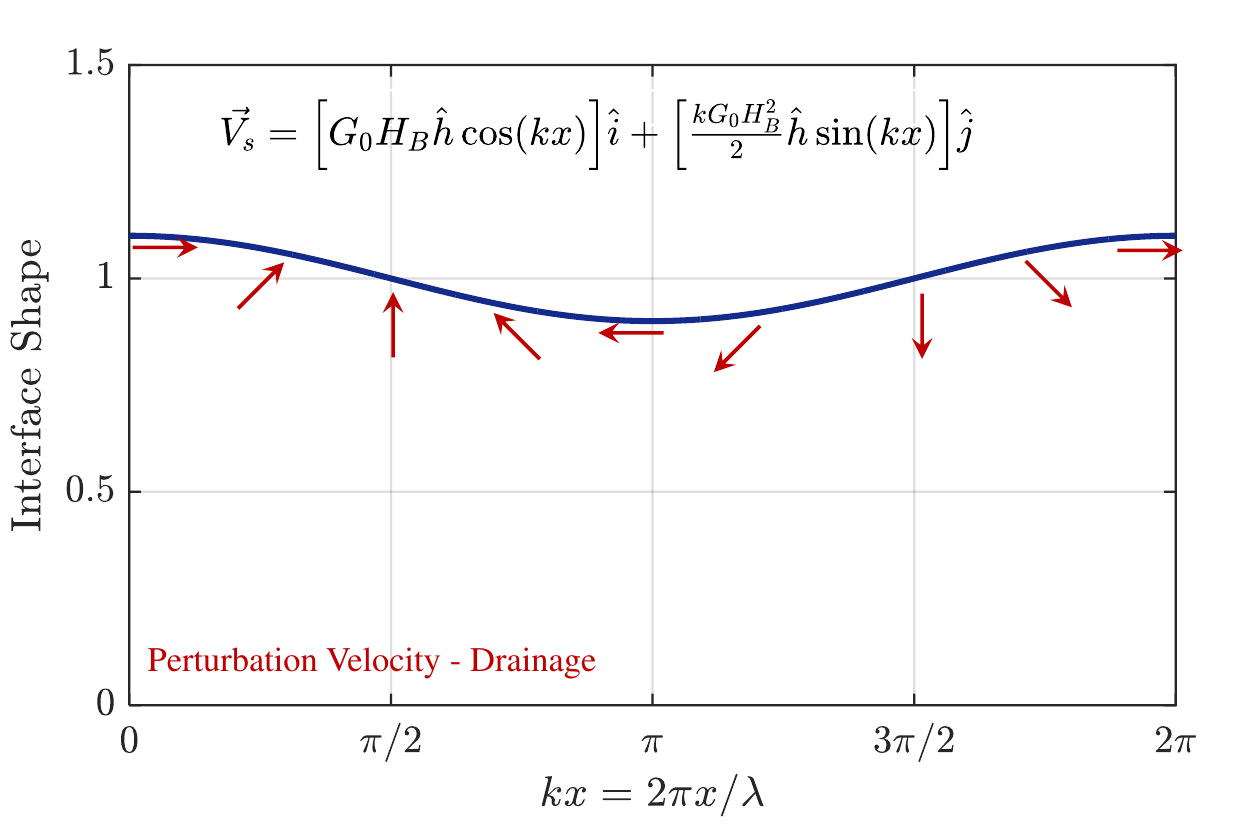}
         \caption{}
         \label{fig:VelocityFields_drainage}
     \end{subfigure}
     \begin{subfigure}[h]{0.45\textwidth}
         \centering
         \includegraphics[width=\textwidth]{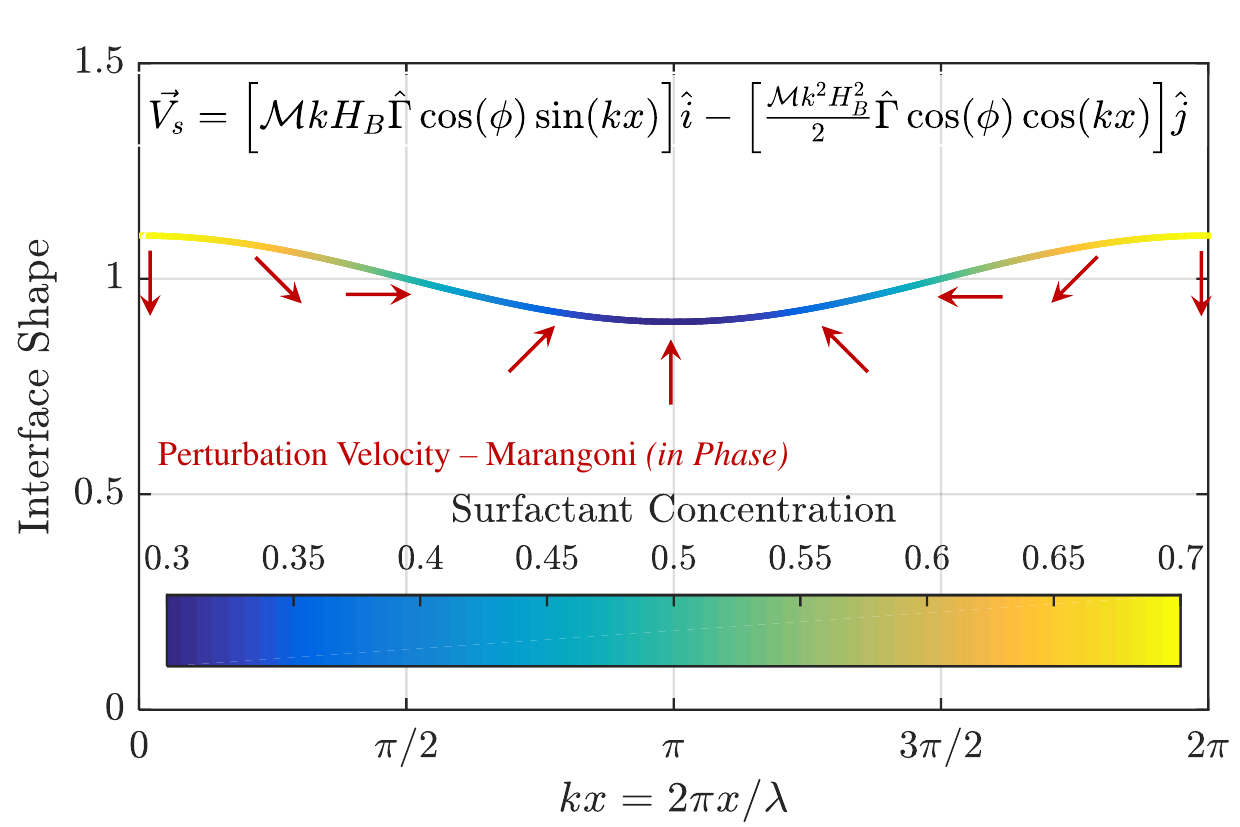}
         \caption{}
         \label{fig:VelocityFields_MarangoniInPhase}
     \end{subfigure}  
    \begin{subfigure}[h]{0.45\textwidth}
         \centering
         \includegraphics[width=\textwidth]{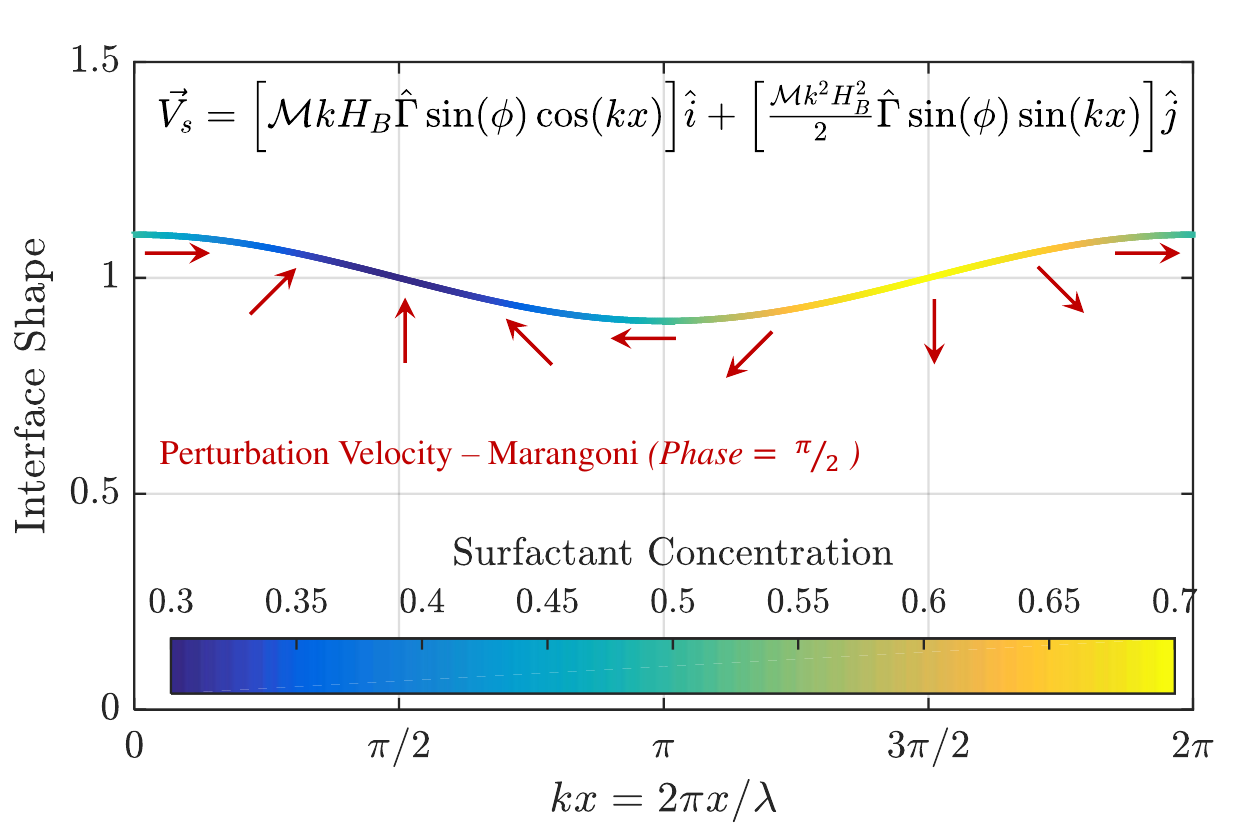}
         \caption{}
         \label{fig:VelocityFields_MarangoniOutPhase}
     \end{subfigure} 
        \caption{ Qualitative representation of perturbation surface velocities generated by various physical pheonomena for a sinusoidal interfacial perturbation and an associated perturbation in surfactant concentration, with a phase difference of $\pi/2$ between the two. Surface velocity field attributed to  (a) van der Waal's dispersion forces, (b) Surface tension, (c) Drainage, (d) and (e) Marangoni convection due to components of surfactant perturbation that is in phase and at a phase difference of $\pi/2$ with the interfacial deflection respectively.}
        \label{fig:VelocityFields}
\end{figure}

To understand the genesis of the stabilizing effect seen in the first mode and destabilization of Mode - 2, we carry out a detailed examination of the surface velocities generated by the two modes. For this, we consider an interface with a sinusoidal deflection imposed on it: $h(x,t) = H_B + \hat{h}\cos(kx)$, along with an perturbed surfactant distribution: $\Gamma(x,t) = \Gamma_B + \hat{\Gamma}\cos(kx+\phi)$.Here, we may arbitrarily specify one of the perturbation amplitudes (say $\hat{h}$). The amplitude ($\hat{\Gamma}$) and the phase shift($\phi$) of the second perturbation is determined by the eigenfunctions of the mode under consideration. In the linear limit, the surface velocities generated by the perturbations as viewed from a frame of reference travelling with the surface speed of the film, can be estimated as:

\begin{gather} 
    V_{s} = \frac{kG_0H_B^2}{2}\hat{h}\sin(kx) + \frac{A}{H_B}k^2\hat{h}\cos(kx) -  \frac{ \sigma H_B^3}{3}k^4\hat{h}\cos(kx) - \Big[ \frac{\mathcal{M}k^2H_B^2}{2}\hat{\Gamma}\cos(\phi)\cos(kx) -\frac{\mathcal{M}k^2H_B^2}{2}\hat{\Gamma}\sin(\phi)\sin(kx) \Big] \label{Eq:TransverseSurfVel}\\
    U_{s} =  G_0H_B\hat{h}\cos(kx) -\frac{3A}{2H_B^2}k\hat{h}\sin(kx) + \frac{\sigma H_B^2}{2}k^3\hat{h}\sin(kx)  + \Big[ \mathcal{M}kH_B\hat{\Gamma}\sin(\phi)\cos(kx) + \mathcal{M}kH_B\hat{\Gamma}\cos(\phi)\sin(kx) \Big] \label{Eq:TransverseSurfVel}
\end{gather}
wherein $V_s$ and $U_s$ denote the surface velocities along the streamwise and transverse directions of the film respectively. Furthermore, the first terms in both the aforementioned expressions represent the perturbation velocity generated by drainage. The second and third terms give the perturbation velocity fields generated by van der Waal's and capillary forces respectively. Finally, the terms within the square bracket are attributable to Marangoni convection. The Marangoni velocity field is separated into two components: one that arises from the surfactant perturbation in phase with the interface deflection: $\hat{\Gamma}\cos(\phi)\cos(kx)$ and another stemming from the surfactant perturbation with a phase difference of $\pi/2$ relative to the interface deflection: $\hat{\Gamma}\sin(\phi)\cos(kx)$. The surface velocities generated by the various physical effects described above is qualitatively depicted in Fig. \ref{fig:VelocityFields} for a phase difference of $\pi/4$ between the surfactant and interfacial perturbation. It may be observed from Figs. \ref{fig:VelocityFields_vdW} and \ref{fig:VelocityFields_Capillary}, that perturbation velocity filed arising from van der Waal's dispersion drives fluid from the troughs to the crests in the films   
and thereby aid in perturbation growth. On the contrary, the capillary velocity field generates a flow from the crests to the troughs which causes a decay of the perturbations. Interestingly, the drainage component in Fig. \ref{fig:VelocityFields_drainage} does neither of the above.  Rather, as seen from the velocity fields, it generates a rightward motion of the surface perturbation. In the case of Marangoni convection, from Fig. \ref{fig:VelocityFields_MarangoniInPhase}, it is evident that the component of surfactant perturbation that is in phase with the interface deflection helps in attenuating the interface deflection. However, as seen from Fig. \ref{fig:VelocityFields_MarangoniOutPhase}, the component of surfactant perturbation that is at a phase difference of $\pi/2$ with the interface disturbance leads only to a rightward motion of the perturbation, with no variation in magnitude. 

It may be noted here that the phase shift $\phi$  is non-zero only for draining films, whereas for a stationary film it is identically zero. From this point and the preceding analysis, it can be observed that the impact of drainage on stability solely manifests through the phase shift and magnitude of the surfactant perturbation that accompanies a given interfacial deflection. For both modes obtained from the linear stability problem, information regarding the aforementioned phase and relative magnitudes is contained within the associated complex eigenfunctions $\tilde{h}e^{ikx}$ and $\tilde{\Gamma}e^{ikx}$. For a given interfacial perturbation, $\tilde{h}e^{ikx}$, marangoni effect is able to stabilize the film better if $\tilde{\Gamma}e^{ikx}$ is large and in phase with $\tilde{h}e^{ikx}$. Conversely, if $\tilde{\Gamma}e^{ikx}$ is small or has a phase difference close $\pi/2$ with respect to the interfacial perturbation, Marangoni convection loses its efficacy in attenuating surface perturbations and consequently, the film becomes less stable.      

Further, we introduce the \textit{projection} function $\zeta_{j}(k)$ given in Eq.~(\ref{Gamma_H_Projection}) as a normalized measure of the component of surfactant perturbation viz. in phase with the interface shape for the $j^{th}$ mode. It may be also reasoned that for a given Marangoni number, the efficacy of surfactant induced stabilization in an eigenmode depends on the corresponding projection function $\zeta_{j}(k)$. A larger projection function implies greater stabilization and vice versa.

\begin{gather}\label{Gamma_H_Projection}
\zeta_{j}(k) = \mathfrak{Re}\Bigg(\frac{\tilde{\Gamma}(k)}{\tilde{h}(k)} \Bigg)
\end{gather}

The key aspects of the linear stability results presented in Fig. \ref{fig:EffG0_GrowthRate_Comp} may now be explained based on the projection of eigenfunctions for the two modes. 
Initially  we shall consider the first mode and the corresponding projection function $\zeta_{1}(k)$ in Fig. \ref{fig:EffG0_Projection_Mode1}. 
In the case of a stationary film and at small drainage, $\zeta_{1}(k)$ is monotonically decreasing with $k$. 
As the critical value of $G_0 = 1.225$ is approached, the projection function experiences a sharp increase at $k_{c,low}$. Subsequently, for higher values of $G_0$, the projection function flattens out.
This indicates an increase in surfactant perturbation and Marangoni stabilization, which consequently accounts for the observed suppression of the growth rate at $k_{c,low}$ (Fig. \ref{fig:EffG0_GrowthRate_Comp}). 
The dispersion relations shows a rapid attenuation of the interface mode  beyond $k_{c,low}$, for values of $G_0>1.225$. 
This can be attributed to the enhanced Marangoni stabilization associated with nearly flat projection function for this mode at higher $G_0$. 
%{\RD{This can be interpreted as follows- van der Waals interaction dominates over the effect of capillarity leading to rupture. Increasing gravitational effect redistributes surfactants/liquid from the bulges to the troughs of the deformed interface thereby aiding the effect of Marangoni stabilization. Hence, drainage suppresses the disturbance growth rate for the first mode.
%What is the genesis of the second mode? As $G_0$ increases, gravity results in deformation of the film with bulges and shallow troughs (think of the shape of a pinned droplet on an incline). This time the surfactant concentration is lower, and hence surface tension is higher, at the bulges.This further draws liquid out of the shallow troughs leading to rupture. This is unique in the sense that Marangoni effect actually drives the instability.}}
Next,  for the second mode, up to the critical value of $G_0$, an increase in drainage leads to a reduction in the projection function $\zeta_2(k)$, resulting in a decrease in the stability of the film. 
Beyond $G_0 = 1.225$, a minor increase in the projection function is observed between the upper and lower cut-off wavenumbers, which explains the stabilizing effect of drainage on the surfactant mode, at higher $G_0$ values.\\

Although perturbation wave speeds do not play a role in determining the stability of the film, a few interesting features regarding the same are worth mentioning. 
In the case of a draining film laden with passive species, Eqs.~(\ref{Eq:SurfactantModeGrowthRate}) and (\ref{Eq:InterfaceModeGrowthRate}) revealed that the surfactant mode perturbations travel along with the film's surface, while the interface mode disturbances propagate in the streamwise direction at twice the speed of the film's surface. 
%The excursion of the wave speeds for both modes from these values can be attributed to the surface velocities generated by Marangoni convection. 
For disturbances of exceedingly large wavelengths, the surfactant perturbation is stretched out over very long scales, thereby making the surface tension gradients infinitesimally small. Consequently, in the long wavelength limit, Marangoni convection becomes negligible, and the wave speeds for both the interface and surfactant modes become identical to those for a film laden with passive contaminants. 
Finally, for rapidly draining films (e.g. $ G_0 = 3$), the wave speeds for the interface and surfactant modes become nearly independent of of the wavenumber and approach their respective long-wavelength limits. This can be attributed to the reduction in the relative magnitude of surface velocities generated through Marangoni convection, with respect to the base state speed of the film's surface. 

\subsection{Effect of Marangoni convection on stability of a draining film \label{sec:EffectofSurfactants}}
\begin{figure}[h]
     \begin{subfigure}[h]{0.42\textwidth}
         \centering
         \includegraphics[width=\textwidth]{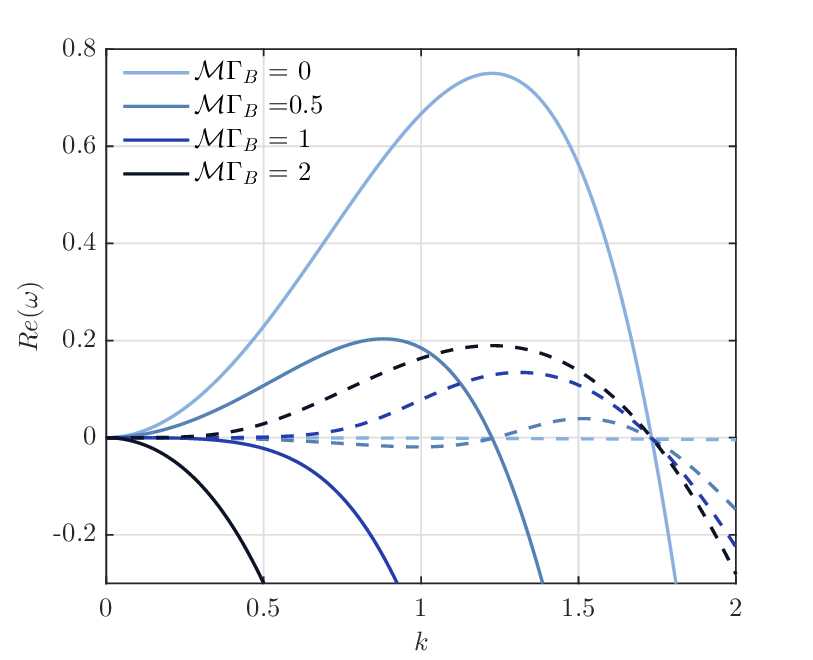}
         \caption{}
         \label{fig:EffMi_GrowthRate_Comp}
     \end{subfigure}
     \begin{subfigure}[h]{0.42\textwidth}
         \centering
         \includegraphics[width=\textwidth]{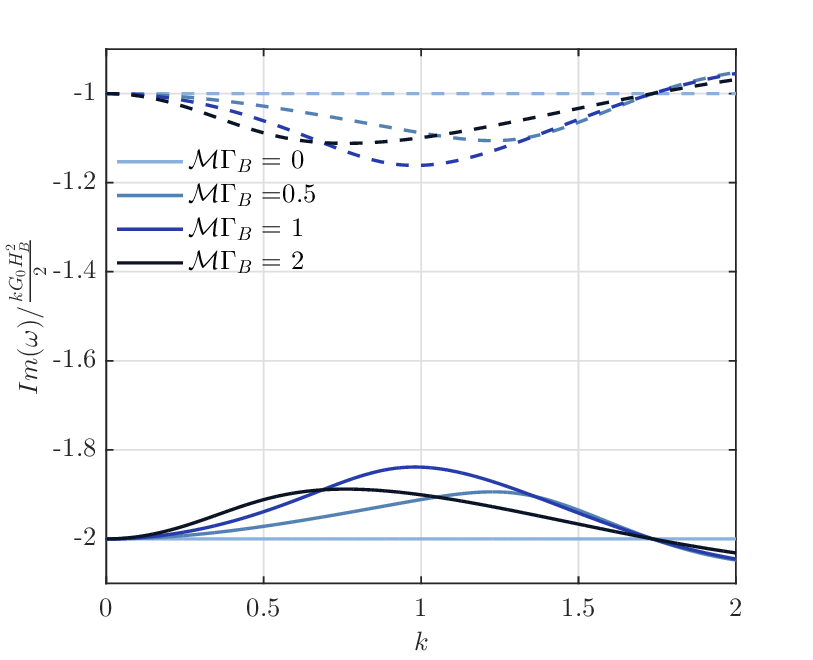}
         \caption{}
         \label{fig:EffMi_Wavespeed_Comp}
     \end{subfigure}
     \begin{subfigure}[h]{0.42\textwidth}
         \centering
         \includegraphics[width=\textwidth]{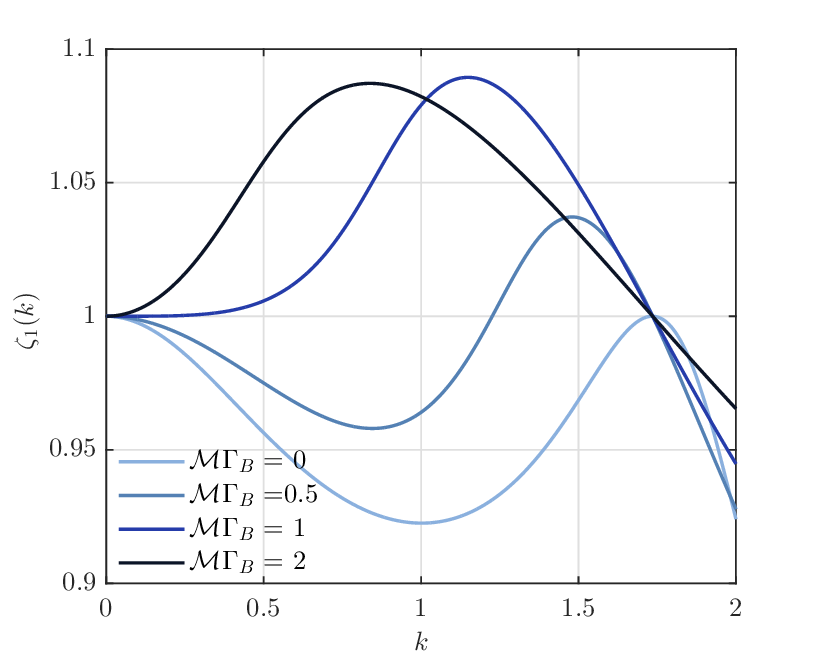}
         \caption{}
         \label{fig:EffMi_Projection_Mode1}
     \end{subfigure}
     \begin{subfigure}[h]{0.42\textwidth}
         \centering
         \includegraphics[width=\textwidth]{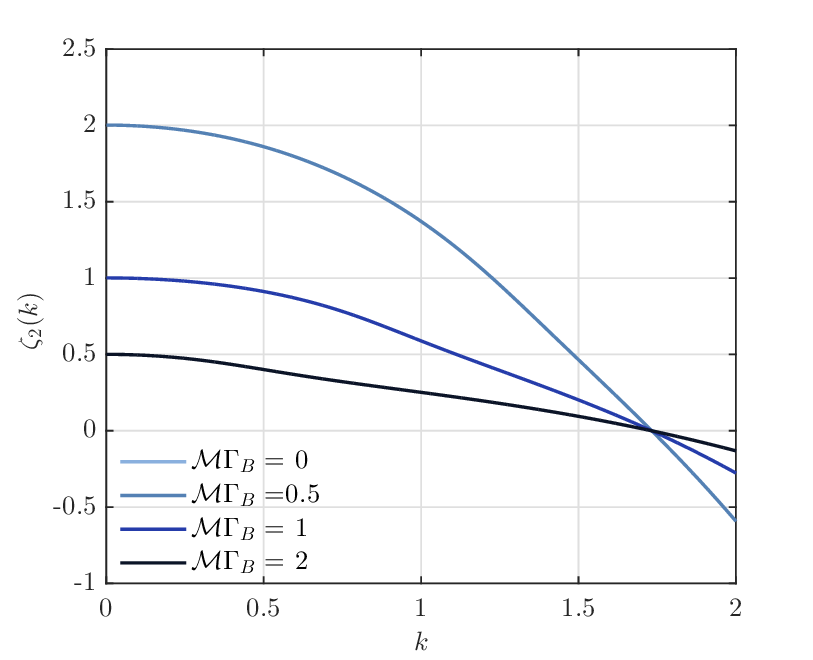}
         \caption{}
         \label{fig:EffMi_Projection_Mode2}
     \end{subfigure}     
        \caption{ Effect of surfactants on stability of the film for the parameters:  $H_B = 1, G_0 = 2,  \sigma = 1, A=1, Pe=1000 $.  For different values of the product $\mathcal{M}\Gamma_B$, we present the following: (a) Non-dimensional perturbation growth rates. (b) Normalized perturbation wave speeds. (c) Projection function for interface mode $\zeta_{1}(k)$. (d) Projection function for surfactant mode $\zeta_{2}(k)$. Solid lines and dashed lines in (a) and(b) denote interface mode (\textit{mode 1}) and surfactant mode (\textit{mode 2}) respectively. It may be noted that for $\mathcal{M}\Gamma_B = 0$, the projection function $\zeta_{2}(k)$ is infinitely large and hence not depicted in the figure.}
        \label{fig:EffectOfMi}
\end{figure}
                                
To assess the influence of surfactants on the stability of a draining film, we compare the dispersion relations for various values of the product $\mathcal{M}\Gamma_B$ in Fig. \ref{fig:EffectOfMi}. 
The reason behind selecting this specific parameter stems from the observation that in the dispersion relation, both $\mathcal{M}$ and $\Gamma_B$ consistently emerge as products, never separately. 
As seen from Fig. \ref{fig:EffMi_Wavespeed_Comp}, in the absence of Marangoni convection, we obtain a single unstable mode, which may be identified as the first mode. 
Upon increasing $\mathcal{M}\Gamma_B$ to $0.5$, we obtain two unstable modes and two cut-off wavenumbers as seen in \S \ref{subsec:EffectofG0}. 
At even higher values of $\mathcal{M}\Gamma_B$, the criteria in Eq.~(\ref{Eq:LowerCutoffWavenumberCriteria}) is violated resulting in the presence of only one unstable mode and a single cutoff wavenumber. 
In this scenario, the instability is entirely attributed to the second mode which is the only unstable mode. 
The aforesaid phenomena may be explained using the reduction in the projection function $\zeta_2(k)$ for the surfactant mode with increase in the surfactant parameter $\mathcal{M}\Gamma_B$, as demonstrated in Fig. \ref{fig:EffMi_Projection_Mode2}. 
%{\RD{Marangoni convection suppresses the instability due to the first mode, but aids the instability due to the second mode as discussed above- you should just focus on this}}. 
This effect is in sharp contrast with that for a stationary film, where the enhanced Marangoni effect is consistently accompanied by a reduction in disturbance growth rates. \\

\begin{figure}[t]
     \centering
              \includegraphics[width=0.6\textwidth]{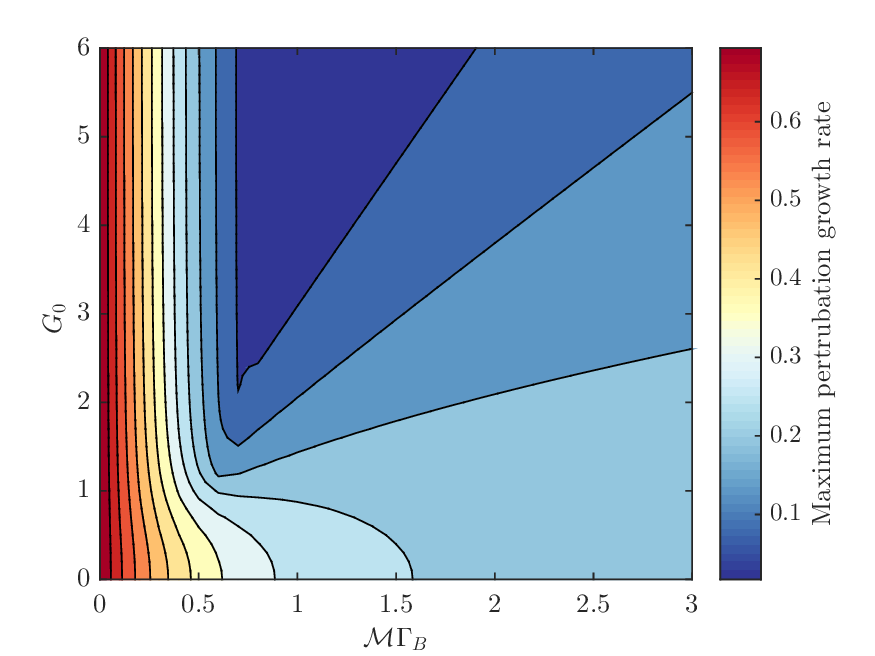}
        \caption{ Contours of maximum disturbance growth rate among all modes for the parameter set: $H_B=1, \sigma = 1, A=1, Pe=1000 $. }
        \label{fig:GrowthRateContour}
\end{figure}

The current observation also raises the question: with all other parameters held constant, which combination of drainage and Marangoni effect renders the film most stable? In Fig. \ref{fig:GrowthRateContour} , a contour plot of the maximum growth rate among all modes of the film as a function of $G_0$ and $\mathcal{M}\Gamma_B$ is presented. The figure reveals that, for all nonzero values of the surfactant parameter $\mathcal{M}\Gamma_B$, the perturbation growth rate consistently decreases with an increase in drainage. However, elevating the surfactant parameter initially stabilizes the film, though at exceedingly high values of $\mathcal{M}\Gamma_B$, it may lead to faster perturbation growth. It may further be seen that rapidly flowing films with intermediate levels of Marangoni convection, denoted by large $G_0$ and $\mathcal{M}\Gamma_B$ around unity, exhibit the highest stability.  
        
 % A second feature of interest is the invariance of the wave speed for the neutrally stable mode at % the the upper cutoff wavenumber. This observation is intriguing because despite the change in the % mode to which the upper cutoff wavenumber belongs, the wave speed remains preserved. If the lower % cut off wavenumber exists, the upper cutoff wavenumber becomes part the surfactant mode. On the % contrary, if the lower cut off wavenumber is nonexistent, the upper cutoff wavenumber belongs to % the interface mode. Nevertheless, at the upper cutoff wavenumber, the neutrally stable mode
 % travels with the same velocity as the surface of the undeflected interface $c = -G_0\frac{H_B^2} 
 % {2}$.
\subsection{Effect of surfactant diffusivity on stability}
\begin{figure}[t]
     \begin{subfigure}[h]{0.45\textwidth}
         \centering
         \includegraphics[width=\textwidth]{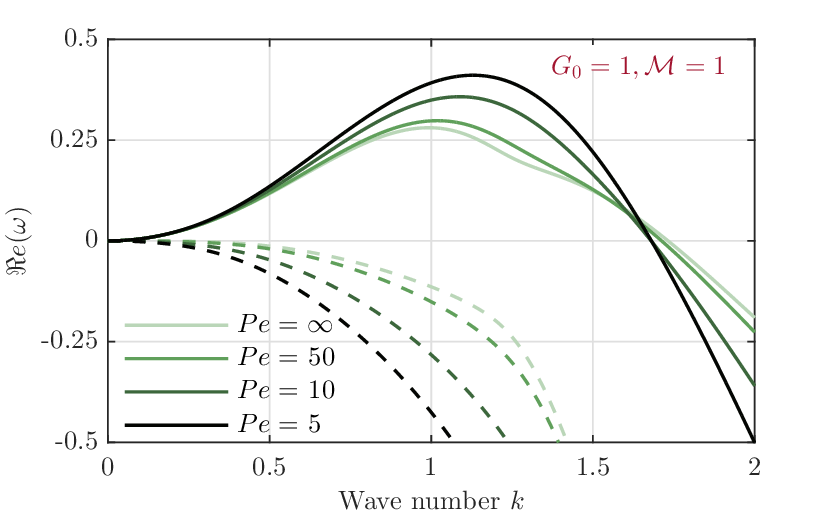}
         \caption{}
         \label{fig:EffectOfPe_LowDrainage}
     \end{subfigure}
     \begin{subfigure}[h]{0.45\textwidth}
         \centering
         \includegraphics[width=\textwidth]{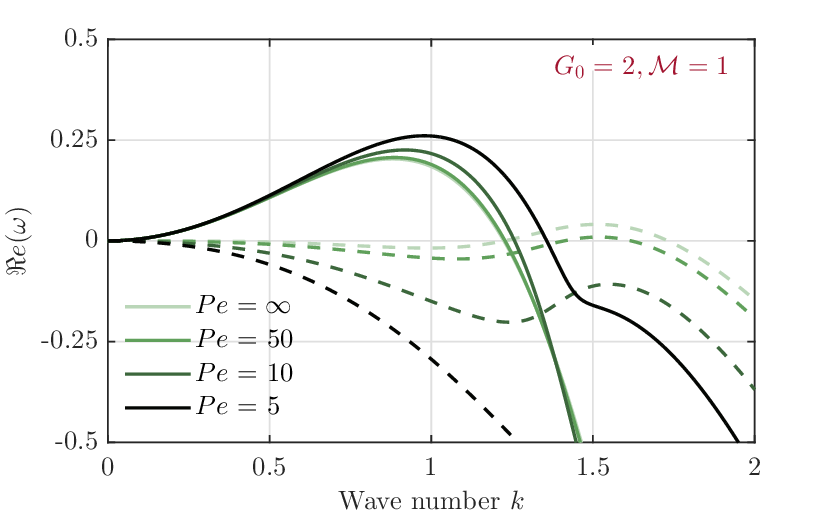}
         \caption{}
         \label{fig:EffectOfPe_RapidDrainage}
     \end{subfigure}
     \begin{subfigure}[h]{0.45\textwidth}
         \centering
         \includegraphics[width=\textwidth]{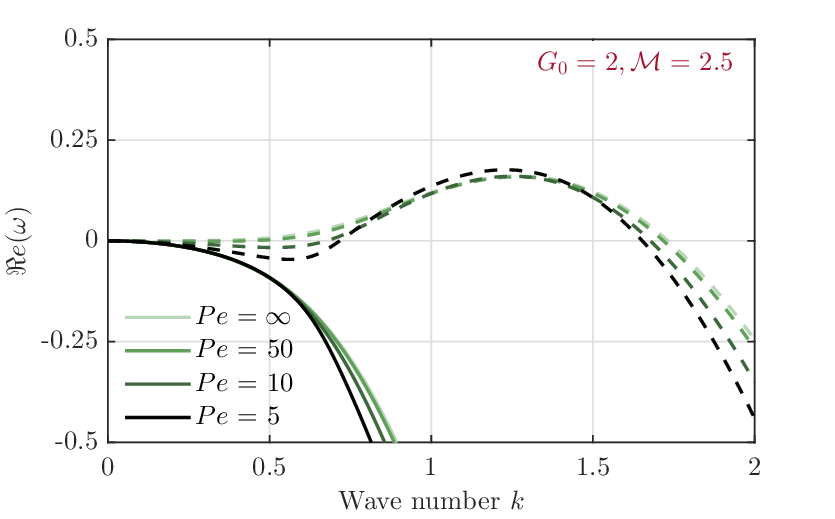}
         \caption{}
         \label{fig:EffectOfPe_SurfDominant}
     \end{subfigure}    
        \caption{Effect of surfactant diffusivity on the stability of the film for three parameter regimes: (a) Low drainage regime with $ G_0 = 1, \mathcal{M} = 1$. (b) Rapid drainage regime with $ G_0 = 2, \mathcal{M} = 1$. (c) Surfactant dominated regime with $ G_0 = 2, \mathcal{M} = 2.5$. Other parameters for all three cases are: $ H_B = 1,  \sigma = 1, A=1, \Gamma_B = 0.5$. Solid lines and dashed lines denote mode 1  and mode 2 respectively.}
        \label{fig:EffectOfPe}
\end{figure}

Note that most of the linear analysis in the preceding sections are carried out in the limit of negligible surfactant diffusion. 
Here, we briefly discuss the influence of surfactant diffusivity on the linear stability. 
The effect of P\'eclet number on the perturbation growth rates for a draining film for three different parameter regimes, namely a low drainage regime, a rapid drainage regime, and a surfactant dominated regime, are shown in Fig. \ref{fig:EffectOfPe}. Before going into the results, it is noted that there are two mechanisms through which surfactant diffusivity can affect the stability of the film. Primarily, for a given interfacial deflection and the associated surfactant perturbation, the higher surfactant diffusivity, as indicated by a smaller P\'eclet number, tends to even out the surfactant gradients, thereby counteracting the stabilizing effect of Marangoni convections \cite{zhang2003slip}. A second scenario may be considered wherein a perturbation in the uniform surfactant concentration generates an unstable interfacial perturbation through the Marangoni effect. In this case, a higher surfactant diffusivity can undermine the Marangoni convection, leading to a reduced interfacial disturbance and consequently an improvement in stability. 
In Figs. \ref{fig:EffectOfPe_LowDrainage} to \ref{fig:EffectOfPe_SurfDominant} the maximum perturbation growth rate for the dominant unstable mode increases with decreasing P\'eclet number, which can be attributed to the former effect, wherein enhanced diffusion of surfactants suppresses the Marangoni convection. Furthermore, in all three figures, the apparent stabilization of the sub-dominant modes with an increase in surfactant diffusivity may be caused by the latter phenomena. 
 Another interesting matter worth exploring is the effect of P\'eclet number on the cut-off wavenumbers. 
 However, unlike in the case of a large P\'eclet number limit, it may not be feasible to obtain simple closed-form expressions for cut-off wavenumbers in the case of films with non-negligible diffusion. Rather, we, attempt to derive approximate solutions for the cut-off wavenumbers using perturbation techniques. For this, we use the perturbative expansion $k = k_{0} + (1/Pe)k_{1} + O(1/Pe^2)$ and $\omega = \omega_{0} + (1/Pe)\omega_{1} + O(1/Pe^2)$. Substituting the expansion into Eq.~(\ref{DispersionRelation}) and imposing the conditions  $\Re e(\omega_{0}) = \Re e(\omega_{1}) = 0 $ and
$ \Im m(\omega_{0}), \Im m(\omega_{1})  \in \mathbb{R}$ as done earlier for the vanishing diffusion limit. This analysis yields the following three approximate solutions for the cut-off wavenumbers:           

\begin{align} 
    k_{c1} \approx k_{c,low} + \frac{6}{Pe} \Bigg( \frac{\Big[k_{c,low}G_0H_B^2/4 - 1/2 \sqrt{(k_{c,low}^2G_0^2H_B^4/4) - (\mathcal{M}k_{c,low}^2\Gamma_BH_B)^2}\Big]^2}{(\mathcal{M}\Gamma_B)^2k_{c,low}^5H_B^5\sigma}\Bigg) \label{Eq:FirstLowerCutoffWaveNumber}
     \end{align}
\begin{align}    
    k_{c2} \approx k_{c,low} + \frac{6}{Pe} \Bigg( \frac{\Big[k_{c,low}G_0H_B^2/4 + 1/2 \sqrt{(k_{c,low}^2G_0^2H_B^4/4) - (\mathcal{M}k_{c,low}^2\Gamma_BH_B)^2}\Big]^2}{(\mathcal{M}\Gamma_B)^2k_{c,low}^5H_B^5\sigma}\Bigg) \label{Eq:SecondLowerCutoffWaveNumber}
\end{align}
\begin{align}
    k_{c3} \approx k_{c,high} - \frac{1}{Pe}\Bigg(\frac{3G_0}{2\sigma(\mathcal{M}\Gamma_B)^2k_{c,high}^3H_B}\Bigg) \label{Eq:NewUpperCutoffWaveNumber}
\end{align}

Wherein, $k_{c,low}$ and $k_{c,high}$ are given by Eq.~(\ref{Eq:LowerCutoffWaveNumber}) and ~(\ref{Eq:UpperCutoffWaveNumber}) respectively. These solutions exist if the criterion given in Eq.~(\ref{Eq:LowerCutoffWavenumberCriteria}) is satisfied. Judging by the sign and relative magnitudes of the first order corrections for cut-off wavenumber given in Eqs.~(\ref{Eq:FirstLowerCutoffWaveNumber}) to ~(\ref{Eq:NewUpperCutoffWaveNumber}), it can be seen that $ k_{c,low} < k_{c1} < k_{c2} \leq k_{c3} < k_{c,high}$. The same is evident from Fig. \ref{fig:EffectOfPe_RapidDrainage}, which denotes the parameter regime that meets the criteria (Eq.~ (\ref{Eq:LowerCutoffWavenumberCriteria})) for co-existence of all three cut-off wavenumbers. As seen from the figure, with an increase in surfactant diffusivity, the second and third cut-off wavenumbers move closer to one another. This is also accompanied by a reduction in the perturbation growth rates for the sub-dominant mode-2 as noted earlier. As the P\'eclet number is continuously further increased, the second mode is completely stabilized and the two cut-off wavenumbers $K_{c2}$ and $K_{c3}$ cease to exist.                  

\section{Nonlinear Analysis\label{sec:NLA} }
In this section, we study the evolution of the film through numerical solutions of the coupled system of nonlinear partial differential equations described in Eqs.~(\ref{Eq:Thickness_Evolution}) and (\ref{Eq:Surfactant_Evolution}). To start computations, the following initial conditions are prescribed for the interface shape and surfactant concentration: 
\begin{gather} 
   h(x,0) = h_B + \mathfrak{Re}\Big(\tilde{h}e^{ik_{m}x}\Big) \label{Eq:InterfaceICs}\\
   \Gamma(x,0) = \Gamma_B + \mathfrak{Re}\Big(\tilde{\Gamma}e^{ik_{m}x}\Big) \label{Eq:SurfactantICs}
\end{gather}
Here, $k_m$ represents the wavenumber associated with the fastest-growing mode obtained from the linear stability analysis for the selected set of simulation parameters. In all cases, the initial interface perturbation is specified as $\tilde{h} = 0.01$. The initial surfactant perturbation $\tilde{\Gamma}$ is obtained by evaluating the eigenfunctions of the dominant mode using Eq.~(\ref{Eq:LSAEigenValueProblem}) at $k_m$ and specifying $\tilde{h} = 0.01$. The governing equations are solved on a computational domain given by $0 \leq x \leq \lambda_m$, with periodic boundary conditions, wherein $\lambda_m = 2\pi/k_m$. Numerical solutions are obtained using an in-house MATLAB solver that uses the central difference method to evaluate spatial derivatives, with the Adam-Moulton scheme for temporal integration. The evolution of the film is tracked till the cusp at the rupture point becomes excessively sharp, rendering the precise calculation of spatial derivatives impractical.

The aforesaid method was validated using results from Burelbach \textit{et al.} \cite{burelbach1988nonlinear} in a previous study \cite{dey2019model} by the present authors. Nevertheless in \S \ref{sec:NLA_FilmEvolution}, a comparison is drawn between the linear and non-linear evolution of the film. A close agreement is observed between the two during the initial stages of perturbation growth, affirming the validity of the current solver. 
\subsection{Thin film evolution \label{sec:NLA_FilmEvolution}}
\begin{figure}[t]
     \begin{subfigure}[h]{0.45\textwidth}
         \centering
         \includegraphics[width=\textwidth]{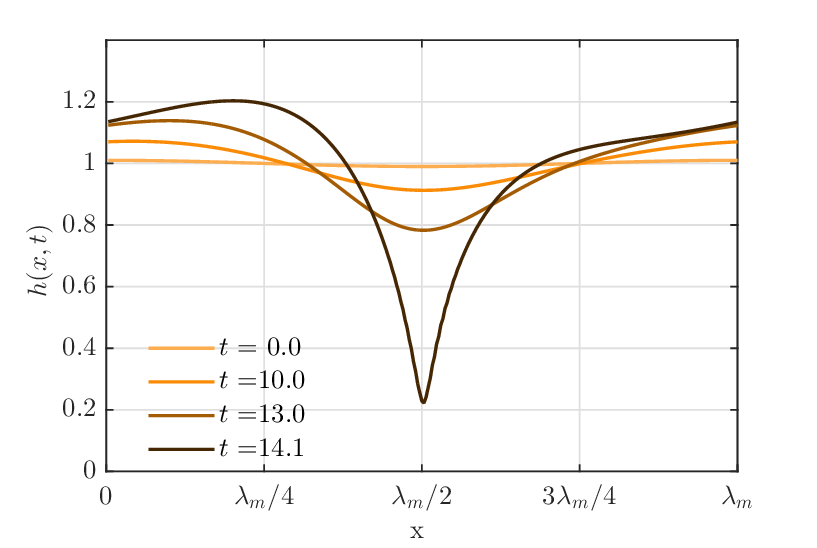}
         \caption{}
         \label{fig:NL_InterfaceEvo_G0=2_Mi=1}
     \end{subfigure}
     \begin{subfigure}[h]{0.45\textwidth}
         \centering
         \includegraphics[width=\textwidth]{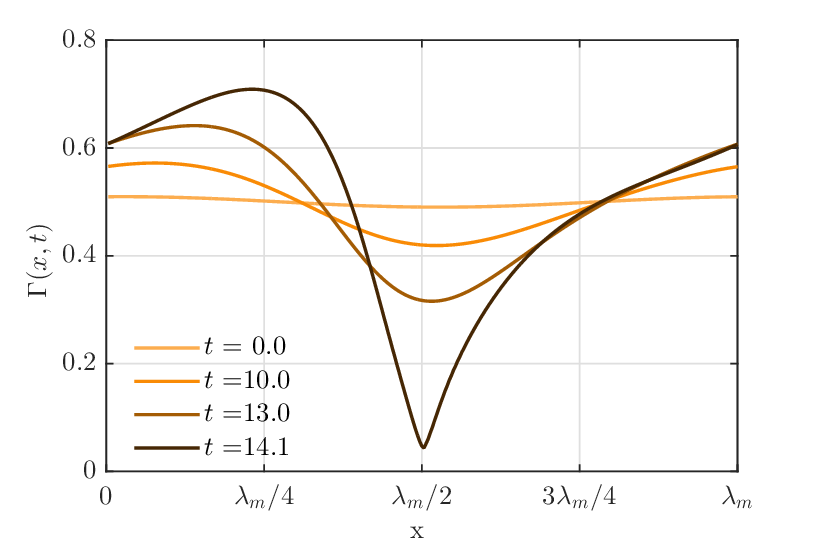}
         \caption{}
         \label{fig:NL_SurfactEvo_G0=2_Mi=1}
     \end{subfigure}
     \begin{subfigure}[h]{0.45\textwidth}
         \centering
         \includegraphics[width=\textwidth]{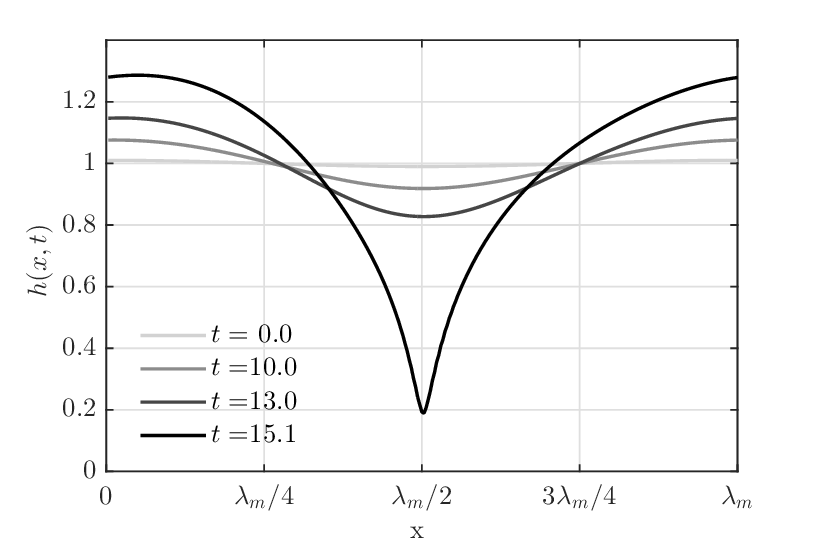}
         \caption{}
         \label{fig:NL_InterfaceEvo_G0=1.35_Mi=2.5}
     \end{subfigure}
     \begin{subfigure}[h]{0.45\textwidth}
         \centering
         \includegraphics[width=\textwidth]{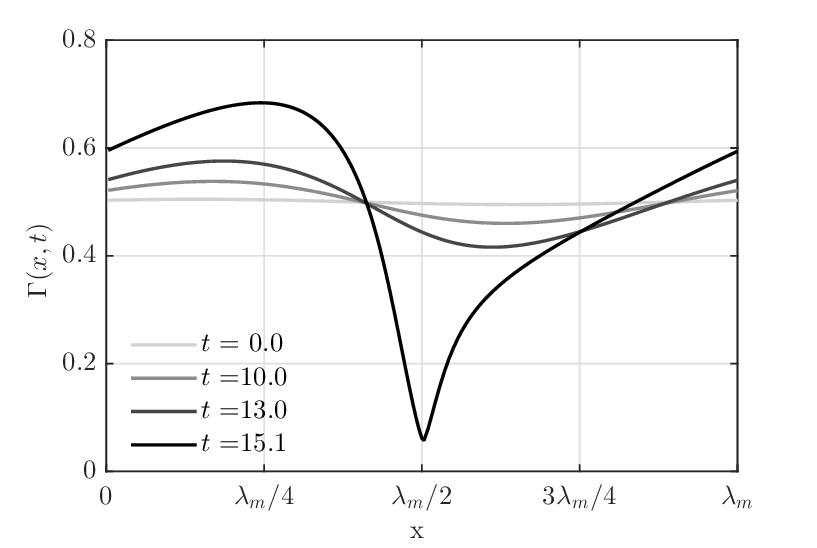}
         \caption{}
         \label{fig:NL_SurfactEvo_G0=1.35_Mi=2.5}
     \end{subfigure}   
          \begin{subfigure}[h]{0.45\textwidth}
         \centering
         \includegraphics[width=\textwidth]{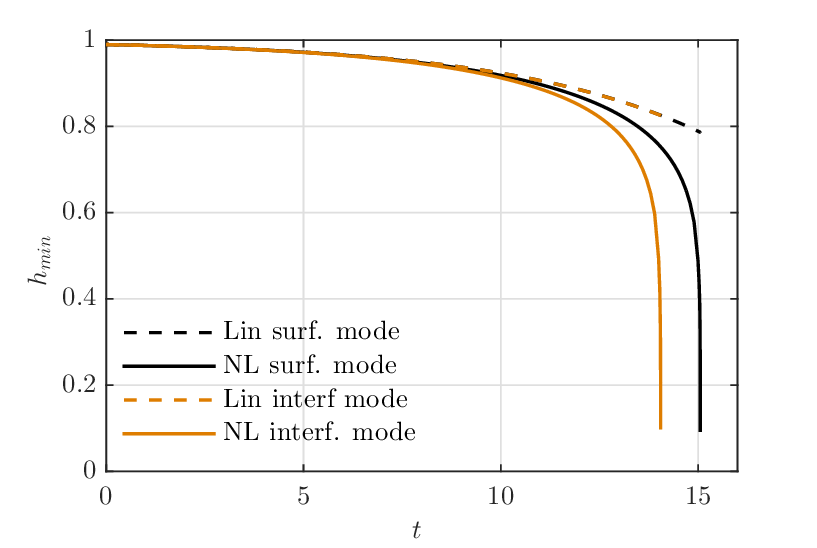}
         \caption{}
         \label{fig:NL_MinHeightEvo_Comparison}
     \end{subfigure}
     \begin{subfigure}[h]{0.45\textwidth}
         \centering
         \includegraphics[width=\textwidth]{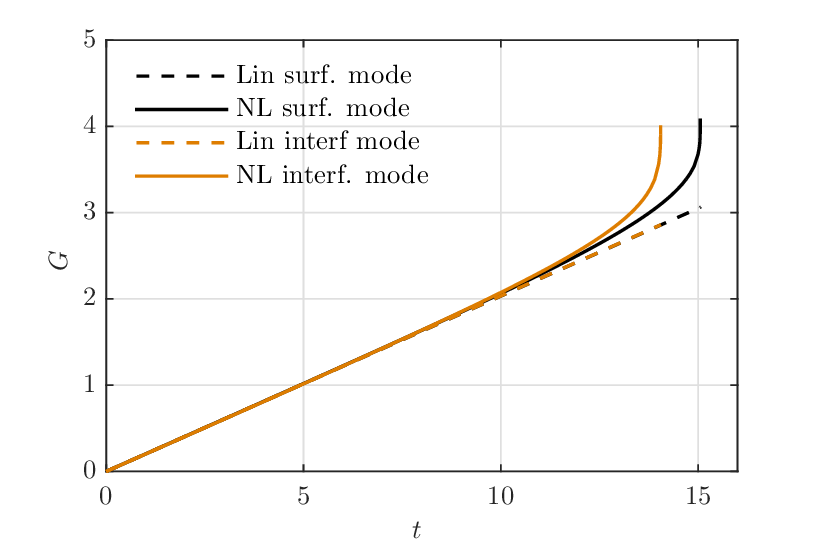}
         \caption{}
         \label{fig:NL_GrowthEvo_Comparison}
     \end{subfigure}   
        \caption{ Nonlinear evolution of thin film for two different parameters set with identical perturbation growth rates ($\mathfrak{Re}(\omega) = 0.203$), but for distinct modes. (a) and (b)  gives the evolution of interface shape and surfactant concentration for film 1 ($H_B = 1,  \Gamma_B=0.5,  G_0 = 2,  \sigma = 1,  A=1,  \mathcal{M} = 1, Pe=1000 $), viz. characterized by a dominant interface mode. (c) and (d) show the evolution of interface shape and surfactant concentration for film 2 ($H_B = 1, \Gamma_B=0.5, G_0 = 1.35, \sigma = 1, A=1, \mathcal{M} = 2.5, Pe=1000$), featuring an unstable surfactant mode. In (e) and (f), the temporal variations of the linear and non-linear estimates for the minimum film thickness and growth metric $G$ are compared for the two cases.}
        \label{fig:NonLinearEvo}
\end{figure}
The numerical results for the evolution of the interface and surfactant concentration, for two sets of parameters, is illustrated in Fig.\ref{fig:NonLinearEvo}.  
These parameters are selected in such a way that they both yield the same perturbation growth rate in the linear limit, but exhibit distinct dominant modes.
For the set $H_B = 1,  \Gamma_B=0.5,  G_0 = 2,  \sigma = 1,  A=1,  \mathcal{M} = 1, Pe=1000  $ the instability is driven by the interface mode, with a maximum perturbation growth rate $Re(\omega_m) \approx 0.203$. 
However, the second group of parameters $H_B = 1, \Gamma_B=0.5, G_0 = 1.35, \sigma = 1, A=1, \mathcal{M} = 2.5, Pe=1000 $ yield a dominant surfactant mode with the same growth rate. 
These two cases shall hitherto be referred to as film-1 and film-2 respectively. It may be noted that the interface and surfactant profile are depicted here as observed from a frame of reference that moves along with the interface perturbations.
From Figs. \ref{fig:NL_InterfaceEvo_G0=2_Mi=1} to \ref{fig:NL_SurfactEvo_G0=1.35_Mi=2.5}, it may be seen that as in the case of stationary films \cite{deWit1994nonlinear,zhang2003slip}, the initial disturbances on the film grows with time, resulting in the formation of a sharp cusp at which the final rupture occurs.
In contrast to stationary films, the interface and surfactant profiles for the draining film are not symmetric about the rupture point. This can be attributed to the cusp's role as a barrier to fluid flow, leading to the accumulation of fluid upstream of the cusp and depletion of fluid downstream from it. 

In Fig. \ref{fig:NL_MinHeightEvo_Comparison}, we present a comparison of the non-linear evolution of the minimum film thickness for film-1 and film-2. Additionally, Fig. \ref{fig:NL_GrowthEvo_Comparison} illustrates a comparison between the linear growth rate $\mathfrak{Re}[\omega(k_m)]t$ and the non-linear growth rate metric \cite{matar2002nonlinear} given by Eq.~(\ref{Eq:NL_GrowthMetric}) for both cases. Results for the two films shown in these graphs indicate that the non-linear results align with the linear estimates during the initial stages of disturbance growth. However, as the perturbations become more pronounced, the non-linear analysis predicts faster growth rates, resulting in a smaller rupture time compared to the linear forecasts. This observation is consistent with previous studies on the rupture of stationary thin films \cite{deWit1994nonlinear,zhang2003slip}.  

\begin{gather}\label{Eq:NL_GrowthMetric}
G = ln \Bigg( \frac{h_{max}(t) - h_{min}(t)}{h_{max}(0) - h_{min}(0)} \Bigg)
\end{gather}

Another feature of interest is that, although both films exhibit identical linear growth rates, the one with the unstable interface mode shows accelerated growth during the later stages of rupture, leading to a quicker rupture.  In the context of the linear studies, the comparable growth rate observed for both films, despite film-2 having a higher Marangoni number than film-1, was attributed to the lower surfactant perturbation projection function $\zeta(k)$ in the former case compared to the latter. A similar observation can be made based on the interface and surfactant profiles of the two films in the early stages on nonlinear rupture. At $t=13$, the interface deflection and surfactant concentration are given in Figs.  \ref{fig:NL_InterfaceEvo_G0=2_Mi=1} and \ref{fig:NL_SurfactEvo_G0=2_Mi=1} for film-1  depict similar profiles, indicating effective Marangoni stabilization. On the contrary, a phase shift is evident between the interface shape and surfactant distribution given in Figs.  \ref{fig:NL_InterfaceEvo_G0=1.35_Mi=2.5} and \ref{fig:NL_SurfactEvo_G0=2_Mi=1} respectively. Following the arguments in \S \ref{subsec:EffectofG0}, this phase shift impedes the Marangoni stabilization during the initial phases of rupture for film-2. However, in advanced stages of rupture, the surfactant redistribution is primarily governed by the interface motion due to van der Waal's forces, and the effect of drainage becomes less pronounced. For instance, at $t=14.0$ for film-1 and $t=15.1$ for film-2, it may be seen that the interface and surfactant profiles become similar to one another. Consequently, surfactant gradients can effectively contribute to the stability of both films. Due to the higher Marangoni number in film-2 compared to film-1, the stabilization effect is more pronounced in the former case. This leads to slower perturbation growth and consequently a higher estimated rupture time for film-2.         

\begin{figure}[t]
     \begin{subfigure}[b]{0.48\textwidth}
         \centering
         \includegraphics[width=\textwidth]{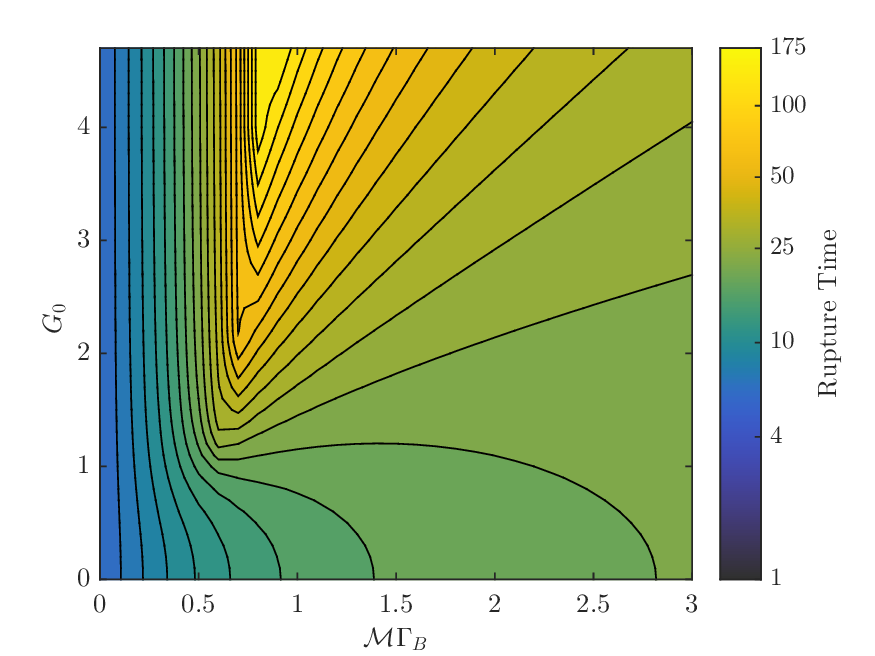}
         \caption{}
         \label{Fig:Linear_RuptureTime_Contours}
     \end{subfigure}
     \begin{subfigure}[b]{0.48\textwidth}
         \centering
         \includegraphics[width=\textwidth]{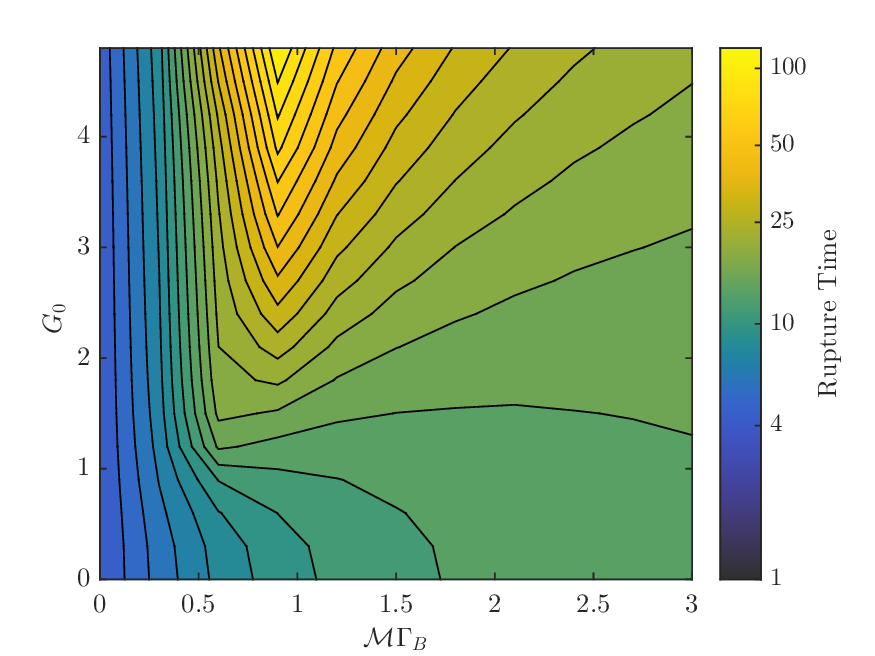}
         \caption{}
         \label{Fig:NonLinear_RuptureTimeContour}
     \end{subfigure}   
    \caption{ Contours of Non-dimensional rupture times for different values of $G_0$ and $\mathcal{M}\Gamma_B$ from (a) Linear stability (b) Nonlinear analysis for the parameter set ($H_B = 1,\sigma = 1,  A=1, Pe=1000 $. For both the estimates, an initial interfacial perturbation of magnitude $\tilde{h} = 0.01$ is considered.  }
    \label{Fig:RuptureTimeContours}
\end{figure}
\begin{comment}
\subsection{Influence of drainage and Marangoni convection on nonlinear stability \label{sec:NonLinear_Contour}}
    
\end{comment}      

The contour plots in Fig. \ref{Fig:RuptureTimeContours} depict variations of linear and nonlinear rupture time estimates with drainage and Marangoni convection. As observed from the plots, the nonlinear rupture times exhibit a similar pattern as the linear estimate, albeit with lower values as noted earlier in \S \ref{sec:NLA_FilmEvolution}.
In the absence of surfactants or when surfactants with a lower Marangoni number are used, drainage has minimal impact on the film's stability. In this regime, the instability is attributable to the first mode, as seen in Fig. \ref{fig:EffectOfG0}. The figures clearly illustrate that films containing surfactants with moderate surface activity levels ($\mathcal{M}\Gamma_B \sim 1$) exhibit greater stability than surfactant-free films.  Furthermore, with increase in drainage,  such films exhibit markedly improved stability owing to enhanced surfactant perturbations.

For stationary films and at low drainage levels, increasing surfactant activity ($\mathcal{M}\Gamma_B > 1$) continues to enhance film stability, with instability primarily driven by first mode. In such films, at higher drainage levels, the second mode becomes the dominant source of instability, as previously discussed in \S \ref{sec:EffectofSurfactants}. Despite the tendency of increased drainage to stabilize films in this scenario, the extent of stabilization remains lower compared to films with $\mathcal{M}\Gamma_B \sim 1$. Consequently, the most stable films, as indicated by their longest rupture times, are observed when high drainage levels are combined with intermediate surface activity levels.       

\section{Conclusions \label{sec:Conc}}

In the present paper, we have studied the concomitant role of gravity and van der Waal's forces in the stability of a draining liquid film laden with insoluble surfactants. Through scaling arguments, we have demonstrated that dispersion forces and gravitational drainage can become simultaneously relevant for a wide range of film thickness and Hamaker constants. Using lubrication theory, we derive a set of coupled nonlinear partial differential equations governing the evolution of the interface and surfactant concentration. A linear stability analysis of the film revealed the presence of two unstable modes, as opposed to a single mode of instability in the case of stationary films. The two modes were identified as an interface mode associated with deflection of the free surface of the film and a surfactant mode triggered by perturbations in the surfactant concentration profiles. In the long wavelength limits, the surfactant mode was observed to travel at the same speed as the surface of the unperturbed film. The interface mode, however, moves at twice the surface speed of the film. We have also shown the existence of two cut-off wavenumbers at which the film becomes neutrally stable. In the limit of vanishing surfactant diffusion, analytical expressions for the cut-off wavenumbers were derived. Furthermore, it was illustrated that the upper cut-off wavenumber reflects an equilibrium between the stabilizing interfacial tension and the destabilizing dispersion forces, as indicated by prior investigations on stationary films. Conversely, the lower cut-off wavenumber represents a unique equilibrium involving van der Waals forces, capillary effects, and Marangoni effects, occurring above a critical threshold of the drainage parameter $G_0$.

Despite the existence of two unstable modes, a linear stability analysis suggests that draining thin films containing insoluble surfactants demonstrate greater stability compared to both stationary films with surfactants and draining films without surfactants. The increased stability can be attributed to the enhanced surfactant perturbation in the interface mode in the presence of drainage.  The lowest perturbation growth rates were observed at large drainage levels and intermediate degree of Marangoni convection ($\mathcal{M}\Gamma_B \sim 1$). At high surfactant concentration and Marangoni numbers, the stabilizing effect is diminished, with the surfactant mode emerging as the dominant mode of instability. 

Nonlinear evolution of the films was studied using numerical techniques and the results show good agreement with the predictions from linear theory, during the initial stages of perturbation growth.  However, nonlinear analysis predicts much shorter rupture times as compared to linear predictions. A comparison of the linear and nonlinear breakup of thin films for different degrees of drainage and Maranagoni convection revealed qualitatively similar rupture times. Nevertheless, the analysis shows that drainage could exert a comparatively stronger stabilizing influence in the nonlinear regime, particularly at lower levels of surfactant activity. Additionally, it was illustrated that nonlinear rupture time in films driven by the interface mode could be slightly shorter than those in films governed by dominant surfactant modes, even though linear studies predict similar perturbation rates.

The results from the present study can potentially have implications for a wide variety of thin films, ranging from precorneal tear films to industrial coatings. Further, it opens up the possibility of maximizing the stability of thin films in industrial settings, through appropriate control of drainage and surfactant dosage. Natural extensions of the current work include investigations into how stability is affected by factors such as wall slip, surface rheology, and the viscoelastic characteristics of the bulk fluid.

\begin{acknowledgments}
 RD and HND acknowledge SERB, DST, Government of India for partial support through grant numbers SRG/2021/000892 and MTR/2019/001158 respectively.
\end{acknowledgments}

\nocite{*}

\bibliography{apssamp}% Produces the bibliography via BibTeX.

\end{document}